\documentclass[galaxies,review,oneauthor,pdftex,10pt,a4paper]{mdpi} 
\firstpage{1} 
\articlenumber{x}
\doinum{10.3390/------}
\pubvolume{xx}
\pubyear{2016}
\copyrightyear{2016}
\externaleditor{Academic Editors: J. Gaite, A. Diaferio}
\history{Received: date; Accepted: date; Published: date}
\pdfoutput=1

 \theoremstyle{mdpi}
 \newcounter{thm}
 \newcounter{ex}
 \newcounter{re}
 \theoremstyle{mdpidefinition}

\newcommand{\vecx}{{\vec x}}
\newcommand{\vecq}{{\vec q}}
\newcommand{\vecS}{{\vec S}}
\newcommand{\vecv}{{\vec v}}
\newcommand{\Msun}{M$_\odot$}
\newcommand{\Msunh}{$h^{-1}$ M$_\odot$}
\newcommand{\Mpch}{$h^{-1}$ Mpc}
\newcommand{\hMpc}{$h$ Mpc$^{-1}$}

\newcommand{\mnras}{{\it MNRAS}}

\Title{Approximate methods for the generation of dark matter halo catalogs
in the age of precision cosmology}

\Author{Pierluigi Monaco$^{1,2,3}$}
\AuthorNames{Pierluigi Monaco}

\address{%
$^{1}$ \quad Universit\`a di Trieste, Dipartimento di Fisica, via Tiepolo 11, 34143 Trieste, Italy; monaco@oats.inaf.it\\
$^{2}$ \quad INAF-Osservatorio Astronomico di Trieste, via Tiepolo 11, 34143 Trieste, Italy; 
$^{3}$ \quad INFN, Sezione di Trieste}

\corres{Correspondence: monaco@oats.inaf.it; Tel.: +39-040-3199131}

\abstract{ Precision cosmology has recently triggered new attention on
  the topic of approximate methods for the clustering of matter on
  large scales, whose foundations date back to the period from late
  '60s to early '90s. Indeed, although the prospect of reaching
  sub-percent accuracy in the measurement of clustering poses a
  challenge even to full N-body simulations, an accurate estimation of
  the covariance matrix of clustering statistics, {not to mention the sampling of parameter space,} requires usage of a
  large number (hundreds in the most favourable cases) of simulated
  (mock) galaxy catalogs. Combination of few N-body simulations with a
  large number of realizations performed with approximate methods
  gives the
  most promising approach to solve these problems with a reasonable
  amount of resources. In this paper I review this topic, starting
  from the foundations of the methods, then going through the
  pioneering efforts of the '90s, and finally presenting the latest
  extensions and a few codes that are now being used in
  present-generation surveys and thoroughly tested to assess their
  performance in the context of future surveys.}

\keyword{Cosmology; Dark matter halos; Large-scale structure of the Universe; N-body simulations}

\begin{document}




\section{Introduction}
\label{section:introduction}

\begin{figure} 
\centering{
\includegraphics[width=15cm]{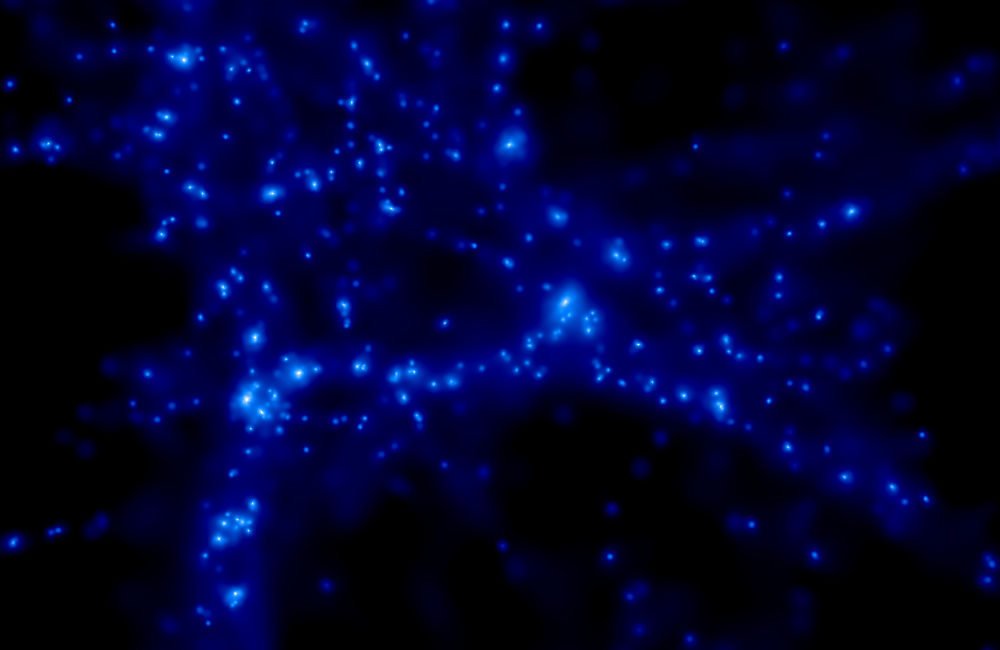}
}
\caption{Large-scale structure in a typical cosmological simulation
  produced by myself with the {\sc GADGET} code \citep{gadget}. The
  box represents a section of a volume of 50 {\Mpch} sampled by $512^3$ particles,
  the color scale gives the projected smoothed density, from black
  (low density) to blue to white (high density). The image was
  produced with {\sc gadgetviewer} (\tt
  http://astro.dur.ac.uk/$\sim$jch/gadgetviewer/index.html).}
\label{fig:simulation}
\end{figure}

The formation of structure in the cosmological $\Lambda$CDM model
(Cold Dark Matter with a cosmological constant $\Lambda$) proceeds
through gravitational evolution and collapse of small fluctuations,
imprinted at very early times during an inflationary period
\citep[e.g.][]{coles2002,mo2010}. These are visible in the cosmic
microwave background as small temperature fluctuations, while they are
still well in the linear regime \citep{planck2014}. Structure in the
dark matter (DM) component, illustrated in
Figure~\ref{fig:simulation}, includes the high-density DM halos, that
contain astrophysical objects (like galaxies and galaxy clusters), and
the low-density, filamentary network that connects them. The small or
vanishing velocity dispersion of CDM particles guarantees that power
is present in fluctuations at all cosmic scales and structure grows in
a hierarchical fashion, smaller perturbations collapsing at earlier
times into less massive halos.

The cosmological model provides initial and boundary conditions to the
problem of the evolution of structure, so, as long as baryons are
treated, to a first approximation, like collisionless DM particles,
the prediction of the growth of perturbations is in principle
determined. 
{ The problem is highly simplified by studying it in the Newtonian limit, on the basis that non-linear structures are well within the horizon\footnote{
 This leaves the question open of how to correctly formulate a Newtonian theory in a general relativistic, inhomogeneous universe;
see, e.g., \cite{buchert2011}
}.}
However, a straightforward, analytic prediction is made
impossible both by the non-linear and by the non-local character of
gravity.
Non linearities make the treatment of the formation of DM
halos untreatable, while simple assumptions like instantaneous
virialization of halos are a clear oversimplification of what is
taking place. Non localities make approaches based on statistical
mechanics, like the BBGKY\footnote{
 The Bogoliubov–Born–Green–Kirkwood–Yvon hierarchy of equations, introduced in cosmology textbooks \cite{coles2002,mo2010}, connects the N-point probability distribution functions of a set of interacting particles, in such a way that the $n$-th equation links the $n$-point with the $(n+1)$-point function. The set of equations cannot be closed without making assumptions on the behaviour of higher-order correlations.
} hierarchy, of little use because closing
the hierarchy requires strong assumptions on the higher-order mass
correlations that are not realistic. Finally, the stochastic character
of fluctuations hampers the straightforward usage of exact solutions
that can be worked out in simplified geometries like planar, spherical
or ellipsoidal.

But this is not, nowadays, a big worry for most cosmologists. The
straightforward, brute-force way out of this problem is that of
sampling a cosmological volume with a large number of particles, and
directly compute their mutual gravitational forces. Several techniques
have been proposed to reduce the computation of forces of a set of $N$
particles from an $N^2$ to a much more manageable $N\log N$ problem,
keeping an acceptable level of accuracy \citep{hockney1981}. Codes are
now able to scale up to thousands of cores, and even better scalings
have been obtained on specific architectures. { Presently,
simulations with $10^{10}$ particles, as the Millennium simulation
\citep{springel2005}, 
are routinely run, while the state of the art has progressed to reach and overtake the $10^{12}$ particle
limit \citep{alimi2012,angulo2012,watson2014,heitmann2015,skillman2014,kim2015,potter2016}.} The experience gathered so far, and the availability of
excellent open source codes, allows any researcher with a minimal
experience in N-body simulations to straightforwardly design, set up
and run a cosmological simulation with $\sim10^9$ particles. With
significant effort, the accuracy with which numerical codes produce
basic quantities like the matter power spectrum or the mass function of
DM halos (for a given halo definition) can reach the percent level
\citep[e.g.][]{heitmann2010,reed2013,schneider2015}. The main limitation for the next
generation of simulations is the difficulty of writing a large number
of full snapshots of all the particles, that are necessary to
synthesize basic predictions at the post-processing level. The
in-lining of post-processing codes, { like halo finders or codes to construct a past light cone,}  can overtake this bottleneck, with
the only limit that the needed final products must be known in
advance, and further detailed analysis may require to re-run the
simulation.

Besides the numerical approach, much effort has been devoted to
understanding the formation of structure with analytic or semi-analytic
methods. In the '90s, a typical paper on this topic might state
that simulations can provide a "blind" solution to the problem, but to
gain insight on what is happening requires a careful comparison of
analytic approximations and simulations. Indeed, much progress in the
understanding of the formation of structure has been gained by
treating simulations as numerical experiments, and analyzing them in
great detail either phenomenologically or by comparing them with
simple models. As a result, though the process of violent relaxation
of DM halos is still poorly understood
\citep{lyndenbell1967,saslaw1985} and the reason why the mass profile
of DM halos takes the double power-law form of \citet{navarro1996}
is still debated, the formation of structure in a hierarchical
Universe is typically regarded as a broadly understood topic. From
this point of view, one would expect approximate methods for the
formation of structure to be out of fashion.

This should be even more true in this age of precision cosmology.
While observations of the cosmic microwave background at $z\sim1100$
with WMAP and Planck have given very accurate constraints on
cosmological parameters \citep{wmap9,planck2015}, the next generation
of surveys of the large-scale structure, like DES\footnote{\tt
  http://www.darkenergysurvey.org/} \citep[Dark Energy
  Survey][]{frieman2013}, eBOSS\footnote{\tt
  http://www.sdss.org/surveys/eboss/} \citep[Extended Baryon
  Oscillation Spectroscopic Survey][]{dawson2016}, DESI\footnote{\tt
  http://desi.lbl.gov/} \citep[Dark Energy Spectroscopic
  Instrument][]{levi2013}, LSST\footnote{\tt http://www.lsst.org/}
\citep[Large Synoptic Survey Telescope][]{abell2009},
Euclid\footnote{\tt http://sci.esa.int/euclid/}\citep{laureijs2011},
WFIRST\footnote{\tt http://wfirst.gsfc.nasa.gov/}\citep[Wide-Field
  Infrared Survey Telescope][]{green2012} or the Square Kilometer Array\footnote{\tt
  http://www.skatelescope.org/} surveys, will provide measurements
that are accurate at the sub-percent level. With statistical errors
beaten down by the very large number of galaxies that can be observed
in the past light cone up to redshift $z\sim1-2$, the
quantification of systematics will be the key element for assessing
the errorbar of measurements. In the most optimistic scenario, it will
assess the significance of { newly found discrepancies} between observations and the
$\Lambda$CDM model that will provide clues to the nature of dark
energy or to modifications of gravity beyond general relativity { (or to the amount of cosmological backreaction within general relativity)}. These
highly accurate observations must be matched with predictions that are
as accurate, and that this high level of accuracy is to be reached
with approximate methods is somewhat counter-intuitive.

But in fact, despite the impressive progress of N-body simulations,
modeling a wide galaxy survey is still a major challenge. Sampling the
very important scale of Baryonic Acoustic Oscillations (BAOs), at
$\sim110$ {\Mpch} (or $k\sim0.07$ \hMpc), requires to simulate a
volume at least $\sim10$ times bigger. Creating a continuous and wide
past light cone, without replicating the information many times along
the same line of sight, requires even larger volumes, 4 $h^{-1}$ Gpc
being a typical minimal value. Besides, resolving the halos that host
the faintest galaxies in a survey, that dominate the number and then
the measurements on which cosmological parameter estimation is based,
requires to resolve halos at least as small as $10^{11}$ \Msun. A
proper sampling of these halos requires particle masses at least two
orders of magnitude smaller. This pushes the requirement on
simulations toward $\sim16,000^3$ particles, for a rather non-linear
configuration at low redshift; no such big simulation has ever been
run. This makes the realization of one single, DM-only simulation a
big challenge. But this is only the start of the story, because a
proper estimation of random and systematic errors needs the production
of a very large number of ``mock'' galaxy catalogs, $\sim10,000$
being a typical number for a brute-force approach. This applies of
course to one single cosmological model, each dark energy or modified
gravity theory would require a similar effort. It is clear that, if
one single simulation is a challenge that will be won in the next few
years, thousands of them are simply out of the question.

This has renewed the interest in approximate techniques, able to
produce realizations of a non-linear universe in a much quicker way
than a standard N-body simulation. The aim here is not to provide the
average of predicted quantities but their covariance, to be inserted
into a likelihood to obtain confidence levels for parameters\footnote{
 This is a simplification of the procedure, 
that is correct only if measurements are Gaussian-distributed. For example, for the power spectrum it can be shown that the variance of a Gaussian process has
as inverse-Gamma (or inverse-Wishart) distribution (e.g. \cite{kitaura2008,jasche2010}). This 
important aspect is however beyond the scope of the present review.
}. But the
interest in approximate methods is not limited to this aspect. For
instance, they are used to generate accurate initial conditions for
the N-body simulations that produce minimal transients { (decaying modes due to inaccuracies in the initial conditions)}, and thus
provide maximal accuracy, when the N-body code is started. Approximate
methods are also used by many astronomers that develop semi-analytic
models (SAMs) \cite{baugh2006,benson2010,somerville2015} for the formation and evolution of galaxies, taking
advantage of the flexibility of such methods that make the production
of model predictions much smoother in many cases. Another interesting
application of approximate methods is the production of constrained
realizations of initial conditions that reproduce, for instance, the
local Universe (see Section~\ref{section:recent}); the Bayesian techniques employed to sample the large
parameter space require many forecasts on what the large-scale
structure is expected to be for a given initial configuration, and
this would lead to painfully long computing times if a proper N-body
code were used.

In this paper I review the state of the art of approximate methods for
the generation of non-linear structure in the age of precision
cosmology. I will mostly focus on DM halos, whose interest relies in
the fact that they host galaxies, the main observable quantity after
recombination. The paper is biased toward the topic of galaxy
clustering, predictions of galaxy lensing can in principle benefit from
approximate methods as well, but its requirements make the usage of
these methods more problematic; I will get back to this point in the
concluding remarks.

I will first quickly review, in Section~\ref{section:foundations}, the
foundations of these methods: Eulerian and Lagrangian perturbation
theories, the Press and Schechter approach and its extensions, density
peaks, exact solutions in spherical and ellipsoidal geometries, and
the issue of halo bias. I will then describe, in
Section~\ref{section:nineties}, the pioneering work done in the '90s
to produce accurate realizations of non-linear density fields.
Section~\ref{section:precision} will describe the latest developments
in the age of precision cosmology and the codes that are now being
used or developed to produce mock catalogs of galaxies for clustering
surveys, while Section~\ref{section:comparisons} will present two
comparisons of methods that have been recently performed.
Section~\ref{section:conclusions} will give a closing discussion.

\section{Foundations of approximate methods}
\label{section:foundations}

Most of the topics of this section are treated by dedicated reviews,
if not by university textbooks, like Peebles' seminal book {\em The
  large-scale structure of the universe} \cite{peebles1980},
\citet{coles2002} or Mo, van den Bosch and White \cite{mo2010}. Here I
will concentrate on the few elements that are needed to understand the
potentiality of the methods described later.

\subsection{Perturbation theories}
\label{section:PT}

After recombination, it is possible to treat dark matter as a perfect
fluid characterized by an equation of state of dust, $p=0$. It is customary to
define a density contrast $\delta$ at a comoving position $\vecx$ 
and at a time $t$ as follows:

\begin{equation}
\delta(\vecx,t) = \frac{\rho(\vecx,t) - \bar\rho(t)}{\bar\rho(t)}
\label{eq:delta}
\end{equation}

\noindent
where $\rho(\vecx,t)$ is the matter density and $\bar\rho(t)$ its
cosmic mean at time $t$,
{ taken to be the homogeneous-isotropic solution \citep[a valid approach in Newtonian cosmology, see][]{buchert1997}.  The density contrast $\delta$ is assumed to be a Gaussian random field (a stochastic field for which all N-point distribution functions are Gaussian multi-variates) with zero mean.}
The condition $\rho\ge0$ implies that
$\delta\ge-1$. Evolution of the fluid under its own gravity is
determined by Euler, continuity and Poisson equations. Under the
assumption that perturbations are small, it is possible to neglect
non-linear terms in the equations and obtain a linear solution with a
growing and a decaying mode. The growing mode evolves like:

\begin{equation}
\delta(\vecx,t) = \frac{D(t)}{D(t_i)} \delta(\vecx,t_i)
\label{eq:growing}
\end{equation}

\noindent
The evolution of density contrast is thus self-similar, its value is
simply rescaled, point by point, by the growth factor $D(t)$,
normalized here as $D(t_0)=1$ ($t_0$ being the actual cosmic time). An
important feature of the growing mode is that it allows a direct
connection of density contrast with peculiar velocity. Let's call
$\phi(\vecx)$ the (suitably rescaled) peculiar gravitational potential that
obeys the equation $\delta(\vecx,t)=D(t)\nabla^2_\vecx \phi(\vecx)$.
Then the velocity field of the growing mode can be written as:

\begin{equation}
\vecv = - H(t) a(t) D(t) f(\Omega_m)\nabla_\vecx \phi
\label{eq:velocity}
\end{equation}

\noindent
where $a(t)$ and $H(t)$ are the scale factor and Hubble parameter, and
$f(\Omega_m)= d\ln D/d\ln a$ is typically expressed as a function of
the matter density parameter $\Omega_m(t)$.

Linear evolution of perturbations is relatively trivial: the density
contrast grows in a self-similar way, so that what is over- or
under-dense will become more and more over- or under-dense. Starting
from Gaussian initial conditions, half of the mass has $\delta<0$, so
the extrapolation of linear theory predicts that half of the mass is
always in underdense regions, possibly where $\delta<-1$.

There are two reasons for this rather unphysical prediction. The first
is rather obvious, linear theory is valid as long as $\delta\ll 1$, so
the extrapolation to high values of the density contrast is to be
avoided. The other reason is more subtle. Non-linear terms in the
Euler and continuity equations involve also velocities, and they can
be neglected not only under the condition that $\delta\ll 1$, but also
under the conditions that displacements of particles from their
initial conditions are ``negligible''. What ``negligible'' means must
be specified. Textbooks like \citet{peebles1980} give the
condition that, at time $t$, $vt/R\ll1$, where $v$ is the velocity of
the mass element at time $t$ so that $vt$ is roughly its displacement
from the position at $t=0$, while $R$ is the typical size of the
smallest perturbation that is present in the density field. The
introduction of a scale $R$ is not arbitrary. The variance of the
density contrast, often called mass variance, can be computed from the
power spectrum $P(k)$ as:

\begin{equation}
\sigma^2(t) = D^2(t) \frac{1}{2\pi^2}\int_0^{\infty} P(k) k^2 dk
\label{eq:sigma2}
\end{equation}

\noindent
For a CDM model, the integral has a logarithmic divergence, due to the
fact that at small scales the power spectrum tends to $P(k)\propto
k^{-3}$ and is generally shallower. To avoid that power at very small
scales gives a very high mass variance $\sigma^2$ even at very small
times, the power spectrum must be truncated somewhere. We will get
back to this point in the next Subsection.

The obvious step beyond linear theory is to calculate higher-order
corrections through a standard perturbation approach, where $\delta$
is the small parameter of the perturbation series. This is known as
Eulerian Perturbation Theory (hereafter EPT), and has been reviewed by
\citet{bernardeau2002} (but see \citet{carlson2009} for a more recent
assessment of the accuracy of perturbation theories). EPT, often
called standard perturbation theory, has very nice points of contact
with the perturbation theory in quantum electrodynamics, so some of
the techniques developed to solve Feynman diagrams can be used to
compute higher-order perturbations. In general, the $n$-th order
determines the statistics at $(n+1)$-th order, so if the Probability
Distribution Function (PDF) of the density contrast remains Gaussian
in linear theory (assuming of course Gaussian initial conditions), it
gets a skewness at the second-order and a kurtosis at the third order.

The name ``Eulerian'' in EPT is due to the use of Eulerian approach to
the evolution of the cosmic fluid, where, e.g., $\rho(\vecx,t)$ is the
density of the fluid element that happens to be in the position
$\vecx$ at time $t$. A different approach is based on the Lagrangian
description of fluid dynamics, where dynamical quantities follow fluid
elements along their trajectories. Let's call $\vecq$ the position of
a mass element at time $t=0$. Its position $\vecx(\vecq,t)$ at $t$ can
be written as:

\begin{equation}
\vecx(\vecq,t) = \vecq + \vecS(\vecq,t)
\label{eq:map}
\end{equation}

\noindent
where, in mathematical terms, $\vecS(\vecq,t)$ is a map from the
Lagrangian space $\vecq$ to the Eulerian space $\vecx$ space. This
approach was pioneered by Zeldovich
\cite{zeldovich1970,shandarin1989}, who noticed that an approximation
to fluid motion (hereafter Zeldovich approximation, ZA) is obtained by
assuming that each mass element keeps its velocity ${\vec u} =
d\vecx/dD = \vecv/a\dot{D}$ unchanged (here we use the growth factor
as time variable):

\begin{equation}
\frac{d{\vec u}}{dD} = 0
\label{eq:zel}
\end{equation}

\noindent
The operator $d/dD$ denotes the Lagrangian or convective derivative.
In terms of the map $\vecS$:

\begin{equation}
\vecS(\vecq,t) = -D(t) \nabla_\vecq \phi
\label{eq:mapzel}
\end{equation}

\noindent
Here $\phi$ is the rescaled peculiar gravitational potential
introduced above, as a function of the initial $\vecq$ coordinate, and
$D(t)$ the linear growing mode. This approximation at early times is
entirely consistent with linear theory; the big difference is that
fluid elements here are supposed to move, while in linear theory they
are not. This difference influences much the predictive power of the
two approaches: trying to extrapolate them toward the highly
non-linear regime, linear theory breaks very soon, while ZA remains
roughly accurate as long as the map $\vecS$ is single-valued, or in
other words as long as the evolution of the fluid is laminar (there are no
regions characterized by multiple streams) and particle orbits do not
cross. As an illustration, the density can be computed by recalling
that the Jacobian determinant, $\det (\partial x_i/\partial q_j)$,
gives the relative change of the volume element under the
transformation from $\vecq$ to $\vecx$, so the density contrast can be
written as:

\begin{equation}
1+\delta(\vecq,t) = \frac{1}{\det (\partial x_i/\partial q_j)}
=\frac{1}{|[1-\lambda_1 D(t)][1-\lambda_2 D(t)][1-\lambda_3 D(t)]|}
\label{eq:deltazel}
\end{equation}

\noindent
In the second passage $\lambda_i$ ($i=1,2,3$) are the eigenvalues of
the Hessian of the peculiar potential $\phi$, $\partial^2 \phi /
\partial x_i \partial x_j$. This equation shows that the property
$\delta\ge-1$ remains true at all times, at variance with linear theory.
It also shows that $\delta$ becomes
infinite at times $D(t)=\lambda_i$. This corresponds to the formation
of ``caustics'' in the density field, due to the crossing of particle
orbits, as thoroughly commented in \citet{shandarin1989}. The largest
eigenvalue $\lambda_1$ will determine the first caustic, and will mark
the breaking of the validity of the approximation.

Figure~\ref{fig:lss} shows the ability of several of the methods
described in this paper to reconstruct the large-scale structure of a
cubic box of 1024 {\Mpch} sampled with $1024^3$ particles, where the
power spectrum is given by the $\Lambda$CDM model; here we show a slab
of 128 {\Mpch} of side and 20 {\Mpch} thick. The first top-left panel
gives the results of the standard {\sc gadget} N-body code
\cite{gadget}, the ZA is in the panel below (mid-left panel).
Large-scale structure is neatly recovered, but structures are fuzzy
and the DM halos are not cleanly recognizable.

As well as linear theory is the first order of EPT, the ZA is the
first order of a Lagrangian Perturbation Theory \citep[LPT][]{buchert1989}, where the small
parameter is the displacement of particles (again, compared to some
coherence scale $R$ of the power spectrum). This theory was { developed by Bouchet et al. 
\cite{moutarde1991,bouchet1992,bouchet1995} and Buchert et al.
\cite{buchert1992,buchert1993,buchert1994,buchert1993b}; see also
\citet{catelan1995}.} Development was first done to
second (2LPT) and third (3LPT) order, higher orders will be discussed
below. Reviews can be found in \citet{buchert1996,bouchet1996,ehlers1997}. The way
these approximations can recover the large-scale structure can be seen
in the middle row of Figure~\ref{fig:lss}, where we show the ZA, 2LPT
and 3LPT approximations. At this level, the most visible effect is a
further thickening of the highly non-linear structures at higher
order, their advantage can be judged by analyzing how well the power
spectrum is recovered. We will get back to this point later in
Section~\ref{section:comparisons}.

A notable application of LPT is the creation of initial conditions for
N-body simulations. A Gaussian density field is generated on a regular
grid, then the ZA is used to compute the displacements of particles
from their initial position, at an early enough time so that
displacements are small and multi-streaming is still absent. Many tests
have been devoted to find the optimal initial time
\citep[e.g.][]{crocce2006}; too early a time, beside increasing the
computational time, would increase the weight of round-off errors, but
a late time would give displacements that are too approximate, giving
raise to decaying transients that have an impact on the results. A
better choice is to use 2LPT to generate the initial displacements
\cite{scoccimarro1998,crocce2006}. This technique is now standard in
the generation of initial conditions for N-body simulations.

\subsection{The need for smoothing}
\label{section:smoothing}

Both EPT and LPT are based on the idea that the density field is
smooth on some scale $R$, so that the power spectrum is truncated and
the mass variance finite. In fact, as mentioned above, for a
completely cold DM particle the evolution of perturbations in the
radiation-dominated epoch will force the power spectrum at
recombination to behave asymptotically as $P(k)\propto k^{-3}$. This
implies a very slow, logarithmic divergence of the mass variance
$\sigma^2$, that is easily cured by assuming that particles have a
non-vanishing free streaming scale. But non-linearities will start to
be important soon after matter-radiation equivalence, so the first DM
halos may form at a very high redshift and at very small masses, of
the order of the Earth mass \cite{angulo2016,ishiyama2014}. Such high
non-linearities that develop so early would in principle make any
perturbation approach useless.

This is however no hamper to the applicability of perturbation
theories, and this is due to the fact that the power spectrum $P(k)$
has a decreasing, negative slope at all scales where non-linearities
are important. Indeed, a consequence of the simplicity of linear
evolution (equation~\ref{eq:growing}) is that, in Fourier space,
perturbations at different wavenumbers evolve independently. This
means that, as long as the field is linear, we are allowed to evolve
large scales ignoring what happens at small scales. Unfortunately this
is not true when the field becomes non-linear: different modes couple,
so their evolution cannot be treated independently. Let's imagine to
decompose, at some initial time $t_i$, the density field into a
low-frequency and a high-frequency part: $\delta(\vecx,t_i)=
\delta_{\rm lf}(\vecx,t_i)+ \delta_{\rm hf}(\vecx,t_i)$. The way we
decompose it is the following: we transform the field into the Fourier
space, then to obtain $\delta_{\rm lf}$ we apply a sharp low-pass
filter, removing all high-frequency modes with $k>=k_f$; a
complementary high-pass filter will be used for the high-frequency
term $\delta_{\rm hf}$. If we evolve $\delta_{\rm lf}$ alone, we will
soon create much power at $k>k_f$. This means that, when evolving the
full density field, transfer of power from low to high wavenumbers
will have a deep impact on the evolution of high frequencies. But when
we evolve $\delta_{hf}$ alone into the non-linear regime, its power
spectrum will develop a tail of power at $k<k_f$ only as $P(k)\propto
k^4$ \cite{peebles1980}. This gives a measure of the expected transfer
of power from small to large scales, due to non-linearities. The
decreasing slope of the cosmological power spectrum will thus
guarantee that this transfer of power has a negligible effect on the
larger scales.

Smoothing thus opens the way to the applicability of all approximate
methods: to avoid that non-linearities on small scales hamper our
predictive power on what is happening at large scales, we can just
filter them out. But we must keep in mind that smoothing introduces
one arbitrary scale $R$ and a filter shape that can take different
functional forms.

This argument applies to N-body simulations as well: sampling a field
on a grid with inter-particle distance $d$ is in fact a kind of
smoothing at the scale $R=d$, and this produces a good representation of
the evolution of perturbations as long as non-linearities present at
unresolved scales do not influence much the resolved dynamics.

\subsection{Press and Schechter and its extensions}
\label{section:PS}

Smoothing at different scales is the base upon which
\citet[][hereafter PS]{press1974} in 1974 (anticipated by
\citet{doroshkevich1967} in 1969) performed a first computation of the
expected mass function of DM halos, that turned out to be surprisingly
accurate \cite{efstathiou1988}.

The main assumptions of the PS approach are the following: (i) the
initial density field is Gaussian, (ii) density is computed by
extrapolating linear theory to the non-linear regime\footnote{It is a
  common misconception that PS is based on spherical collapse; this is
  only used to give a plausible value to $\delta_c=1.686$, but in fact
  it is based on extrapolating linear theory.}; (iii) regions with
$\delta>\delta_c$, where $\delta_c$ is a parameter of order unity,
will end up in DM halos; (iv) the density smoothed on a scale $R$,
with a top-hat filter, will give information on halos of mass
$M=\bar\rho\, 4\pi R^3/3$ (up to a constant that may depend on the
filter).
Under these assumptions, the fraction of matter in halos more massive
than $M$ can be equated to the probability that the density contrast,
smoothed on a scale $R$, is larger than $\delta_c$:

\begin{equation}
\frac{1}{\bar\rho}\int_M^\infty M' n(M') dM' = P(>\delta_c) = 
\int_{\delta_c}^\infty \frac{1}{\sqrt{2\pi\sigma^2(R)}} 
\exp\left( - \frac{\delta^2}{2\sigma^2(R)} \right) d\delta 
\label{eq:PS2}
\end{equation}

\noindent
The mass function $n(M)dM$, the number density of halos with mass
between $M$ and $M+dM$, can be readily found by differentiating
equation~\ref{eq:PS2} with respect to the mass.

The big problem with the original formulation of this mass function
theory is that, when $\sigma^2(R)$ goes to infinity, 
{ $\delta_c\ll\sigma(R)$ and 
the right-hand
side of equation~\ref{eq:PS2} becomes an integral from $\sim0$ to $\infty$
of a Gaussian}, that gives $1/2$. This means that for infinite variance
only half of the mass is expected to be found in DM halos. The reason
why this happens is that PS is based on linear theory, that cannot
change the sign of $\delta$, so if half of the mass is in underdensities then
half of the mass will end up in voids. Press and Schechter fixed this
problem by introducing a ``fudge factor'' of 2, but with the excursion
sets approach \cite{epstein1983,peacock1990,bond1991} a very nice and
elegant justification was found. A basic treatment can be found in the
Mo, van den Bosch and White textbook, while reviews of the mass
function and excursion sets theory have been written by 
\citet{monaco1998} and by \citet{zentner2007}. This approach
explicitly deals with the fact that predictions are based on a field
that is smoothed on an arbitrary scale $R$, and that predictions at
different smoothing scales can be in contradiction; this has been
dubbed the cloud-in-cloud problem. It can be demonstrated that the
formulation of a mass function of halos can be recast into computing
the rate at which a set of random walks, of $\delta(R)$ as a function
of $\sigma^2(R)$, upcross an absorbing barrier at $\delta=\delta_c$;
the random walk is a Markov process if the rather extreme ``sharp
k-space'' smoothing filter is used; this is a sharp low-pass filter in
Fourier space, like the one used in the previous Subsection.

Excursion sets theory is the basis for the Extended Press and Schecther
(EPS) approach: a mass element can be part of halo $M_1$ at time $t_1$
and of halo $M_2>M_1$ at time $t_2>t_1$, and this information can be
used to reconstruct the full merger history of a halo. This was first
proposed by \citet{bond1991} and developed by \citet{bower1991} and
\citet{lacey1993}, then it was explored in other papers
\cite{shethlemson1999,somerville1999,vandenbosch2002}. A further
extension is based on the idea that improved predictions can be
obtained using a ``moving barrier'', that amounts to using a function
of $\sigma^2$ as a barrier, in place of the constant $\delta_c$
\cite{sheth1999,sheth2002}. This was originally motivated using
ellipsoidal collapse, that I will describe in next Subsection.

The problem with the excursion sets and EPS approaches is that they
are phenomenological in nature, and they are not guaranteed to give a
reliable approximation of the formation of DM halos. If the agreement
of PS with the first simulations in the mid '80s was surprising, later
works have demonstrated that theory and numerical experiment are in
disagreement. In a proceeding of 1999 \cite{monaco1999} I was
proposing that PS is ``a simple, effective and wrong way to describe
the cosmological mass function''. It is missing so many elements, that
happen to compensate each other, that it works more by the central
limit theorem than by catching the main physics at play. Also, the
construction of consistent of merger trees is rigorously possible only
for a white-noise $P(k)$ \cite{sheth1998}, while in realistic cases a
lot of care is needed to obtain merger histories that, to a good level
of approximation, are self-consistent and accurately reproduce the
results of simulations \citep[e.g.][]{shethlemson1999,somerville1999}.
However, PS and EPS are still widely used for a number of
applications, mostly to have a quick prediction of the halo mass
function and to generate merger trees for SAMs of galaxy formation.

The EPS approach attempts to reconstruct the hierarchical formation of
extended objects, the DM halos, through the statistics of mass
elements. Alternatively, one can start from the statistics of peaks in
the smoothed linear density field, $\delta(\vecq;R)$, assuming that
one peak will collapse into one halo of mass $M\sim\bar\rho R^3$. The
statistics of peaks of a stochastic Gaussian field is a classic topic
of research in mathematics, well described in the textbook of
\citet{adler1981}. Anticipated once more by \citet{doroshkevich1970},
\citet{peacock1985} and especially \citet{bbks} imported this
knowledge into cosmology, and for a few years a large number of papers
were devoted to using and extending the peak formalism to model dark
matter halos in several contexts.

\subsection{Ellipsoidal collapse}
\label{section:ell}

Exact solutions of non-linear evolution of perturbations can be found
only in three cases: planar, spherical and ellipsoidal perturbations.
For planar perturbations, the ZA can be
demonstrated to be the exact solution up to orbit crossing. A
spherical perturbation of radius $R$ can be treated
as an universe with mean density equal to the density within
$R$, so that its evolution is described by the solution of Friedmann
equations up to collapse into a singularity (the equivalent of a big
crunch), at least as long as shells at different radii do not cross.
Shell crossing is indeed a spherically-symmetric form of the orbit
crossing that limits the validity of LPT. In the case of a spherical
top-hat, an easy way to treat the resulting singularity is to assume
that virialization takes place when it forms, either instantaneously
or after a bouncing time. Most uses of spherical collapse assume that
the initial perturbation has a top-hat density profile, described by a
Heavyside step function in radius, and that virialization takes place
immediately at the formation of the singularity.
Spherical collapse has been widely used to understand the formation of
DM halos. The problem with it is that any deviation from spherical
symmetry grows with time, so even if this may be a fair
approximation at early time, all the transients that lead to a
filamentary network before the complete formation of a relaxed halo
are completely missed.

The next level of sophistication, the evolution of a homogeneous
ellipsoid, is still treatable analytically { (this applies to inhomogeneous models as well, see e.g. \citet{kerscher2001})}, though the resulting
integro-differential equations need numerical integration
\cite[e.g.][]{bond1996}.  However, \citet{nadkarni2016} have recently
found a formulation of ellipsoidal collapse that does not involve
integrals. Ellipsoidal collapse was already discussed in
\citet{peebles1980}, and has been used by several authors to improve
the description of the collapse of structures.

Ellipsoidal collapse has an interesting point of contact with the ZA.
The ellipticity of the perturbation grows with time, and this is
because collapse proceeds at a different pace on the three directions, the
shortest axis collapsing fastest. So the peak will first collapse into
a so-called pancake, a flattened structure, then into a filament, then
into a knot.

This approximation has in fact been used to model the non-linear
evolution of both density peaks and generic mass elements. Regarding
peaks, very high peaks tend to have nearly spherical excursion sets,
but a better approximation is clearly that of describing them as
homogeneous ellipsoids, and use ellipsoidal collapse to describe their
evolution. This raises the question on how to define the collapse. It
is quite natural to wait for the collapse on the third axis before
promoting the peak to a halo, the other collapses being transients
that lead from the decoupling from Hubble expansion to the full
virialization. But this requires the further approximation that
virialization takes place separately in the three dimensions at
different times, or that the thickness of the objects along
each collapsed dimension remains frozen after collapse.

Peaks are defined on a given smoothing scale. Changing the smoothing
radius will result in a hierarchy of peaks on different scales that
are nested in a non-trivial way, resembling the hierarchical formation
of the halo through accretion and merger. So the cloud-in-cloud
problem translates here into a peak-in-peak problem. The first to
fully address this problem were \citet{bond1996}, with a
semi-analytic algorithm that identifies peaks and treats their
evolution by numerically integrating the equations of ellipsoidal
collapse, freezing the axes just before their collapse and waiting for
the collapse of the third axis. Recent developments in this direction
will be reviewed later in Section~\ref{section:recent}.

Conversely, ellipsoidal collapse can be used to model the evolution
not of an extended region but of a mass element, as proposed by myself
in 1995 \cite{monaco1995,monaco1997}. Take the peculiar potential
$\phi$ and expand it in a Taylor series:

\begin{equation}
\phi(\vecq) = \phi(\vecq_0) + (q_i-q_{0i}) \frac{\partial\phi}{\partial q_i}(\vecq_0)
 + \frac{1}{2}(q_i-q_{0i}) (q_j-q_{0j}) 
\frac{\partial^2\phi}{\partial q_i\partial q_j} (\vecq_0)
 + ...
\label{eq:taylor}
\end{equation}

\noindent
where summation over repeated indices is assumed.
While the zeroth order is immaterial and the first order determines a
bulk flow of the mass element, the second-order term is the first one
to determine its density evolution. It takes the form of a quadratic
potential, determined by the Hessian of the potential in that point\footnote{ A similar approach is used by \cite{hahn2007} to quantify the topology of large-scale structure.}.
This is the same potential as an extended homogeneous ellipsoid, so
one can use this solution to model the evolution of the mass element. In
fact, one can apply LPT to third order to find a very useful
approximation to the numerical evolution of the ellipsoid \cite{monaco1997}. 

With respect to the evolution of an extended region, the definition of
collapse should be changed. Collapse on the first axis of the
ellipsoid marks orbit crossing, or the transition to the highly
non-linear regime where perturbative approaches do not apply any more.
But first-axis collapse does not imply the formation of a relaxed
object like a halo. Indeed, the filamentary network that joins halos
(Figure~\ref{fig:simulation}) is an example of mass that has collapsed
on { at least one axis but has not suffered violent relaxation}. However, it would be unwise in
this context to wait for collapse of third axis. One reason is that
the extrapolation beyond the first caustic is not safe, but another
reason is that there is a deep difference between the geometry of
collapse of an extended region and of a mass element. As an elementary
example \citep[already presented in][]{monaco1998}, a spherical peak
with a Gaussian density profile will collapse on all directions at the
same time, but it is easy to demonstrate that any of its mass
elements, with the exclusion of the one exactly at the peak (a set of
zero measure), will collapse on two directions only, but not on the
radial direction. Waiting for collapse on the third axis at the mass
element level would result into a halo of vanishing mass, surrounded
by a spherically symmetric distribution of ``filaments''.

\subsection{Halo bias}
\label{section:bias}

Halos, as well as galaxies, are unfaithful, biased tracers of the
underlying density field. The relation between halo and matter density
is a very lively field of research, a necessary first step to better
understand the much more complicated issue of galaxy bias, where
baryon physics determines how halos and sub-halos are populated by
bright objects. The foundations of this topic are well reviewed in
\citet{mo2010} and in \citet{cooray2002}, I will just recall here the
main concepts that are needed to better understand what follows.

The idea behind most attempts to analytically predict halo bias is the
peak-background split, introduced in \citet{bbks}. Assuming that a
halo is produced by the collapse of a peak of the linear density
field, its properties will be determined by scales of order or smaller
than its Lagrangian radius. Larger scales will have the main effect to
modulate the number of peaks higher than a given threshold, thus
producing a correlation between large-scale matter density and number
density of halos. One can then divide (once more) the density field
into a small-scale (high frequency) and a large-scale (low frequency)
part, the former giving the condition for halo formation and the
latter a modulation of their number density.

The first attempt to model bias, before Bardeen et al., was due to 
\citet{kaiser1984}, who computed the correlation function of excursion
sets higher than some high threshold $\delta > \nu\sigma$. On large
scales (for two-point correlation function $\xi\ll 1$) the correlation
function scales linearly with the matter one:

\begin{equation}
\xi_{>\nu} (r) \sim b_1^2 \xi(r) \label{eq:kaiser}
\end{equation}

\noindent
(The expression for $b_1$ can be found in that paper). This is what
happens when halo density is a linear function of matter density:

\begin{equation}
\delta_h(\vecx) = b_1 \delta(\vecx) \label{eq:linearbias}
\end{equation}

\noindent
This was expanded by \citet{bbks}, who computed the
correlation function of peaks of the linear density field, obtaining a
similar result. However, as noticed by \citet{bagla1998}, the
latter work implies an exponential increase of bias for very rare
objects, and this explains why the clustering of luminous
high-redshift galaxies grows with redshift while, going backwards in
time, the clustering of matter decreases.

These works were neglecting the displacements of matter and halos from
their initial, Lagrangian positions, so they were predicting a
``Lagrangian'' bias. The approach based on excursion sets was pushed
forward by \citet{mo1996} and \citet{sheth1999,sheth2001}. They used
the EPS formalism to compute the conditional mass function of halos
that lie in an overdense region; this is a similar problem as finding
the merger tree of a halo: instead of requiring that a mass element is
in a halo of some mass $M_1$ at $t_1$ and of mass $M_2$ at $t_2$, it
is required that, at the same time $t$, a mass element lying in a halo
of mass $M$ is in some overdensity at a larger scale. Then they
defined the (Eulerian) bias as the ratio of the conditional versus
unconditional mass functions, corrected for the displacement of halos
using spherical collapse. This scheme can predict the mass dependence
of halo linear bias as well as the parent EPS scheme can predict the
halo mass function \cite{jing1998}. An improved scheme, based on the
moving barrier formalism, was given by 
\citet{sheth2002}.

But linear bias is just a first approximation of the relation between
halo and matter density. On the assumption that bias can be expressed
as a deterministic function, this can be Taylor-expanded, and a
hierarchy of bias terms can be introduced. The relation between these
and the moments of large-scale correlations was first studied by
\citet{fry1993}. But the relation may deterministically depend on other
quantities like tidal fields, giving rise to ``non-local'' bias
\cite{chan2012,sheth2013}, or it may not be deterministic; bias
stochasticity was addressed by \citet{dekel1999}. This of course needs
a specification of the statistics of stochasticity, and the simplest
assumptions (Poisson, Gaussian) may not be very accurate.

\section{Approximate methods in the '90s}
\label{section:nineties}

A large effort was devoted in the '90s to develop approximate methods
to produce predictions of the clustering of DM halos that could be
compared with data to constrain cosmological parameters, in an age
where N-body simulations were still very costly and numerical
stability was still to be fully demonstrated. These efforts were
thoroughly reviewed by \citet{sahni1995}.

\subsection{Lognormal model}
\label{section:lognormal}

A very simple way to obtain an approximation of a non-linear density
contrast field is that of taking the exponential of a Gaussian field.
Indeed, if the PDF of the density contrast is assumed to be Gaussian
in linear theory, the quantity $\exp(\delta)-1$ is $\simeq\delta$ if
$\delta$ is small, but is always $>-1$ even when $\delta$ becomes very
negative. \citet{coles1991} proposed this lognormal
model as a very simple description of the non-linear evolution of
perturbations.

\subsection{Adhesion theory}
\label{section:adhesion}

Many approximate methods developed in those years were extensions of
the ZA, trying to fix or limit its tendency to spread collapsed
structures. One very elegant way to avoid the thickening of
``pancakes'', the so-called adhesion model of \citet{kofman1988} \cite{gurbatov1989,kofman1992}, is to
add a very small viscosity term in the equations of motion
(Equation~\ref{eq:zel}):

\begin{equation}
\frac{d{\vec u}}{dD} = \nu\nabla^2_{\vecx} \vecx
\label{eq:adhesion}
\end{equation}

\noindent
This is a version of the Burgers' equation, well known in fluid
mechanics. With a very small but non-zero $\nu$, the viscosity term
becomes important at the formation of the caustic, having the effect
of forcing particles to stick to the collapsed web. { The adhesion model and the sticking particle model can in fact be derived from kinetic theory with its exact coefficient \citep{buchert1998}.
Adhesion has not received much recent attention after the '90s, with some exceptions like \citep{menci2002}.}

\subsection{Extensions of ZA}
\label{section:extensions}

Other modifications of the ZA were presented by 
\citet{matarrese1992}, who devised a ``frozen flow'' approximation
based on keeping the velocity field fixed in Eulerian coordinates, so
that matter moves like massless test particles, and the ``linear
potential'' approximation of \citet{bagla1994},
who extrapolated the constancy of the peculiar gravitational
potential, valid in linear theory, to the non-linear regime. One
problem with these approximations is that, while limiting but not
solving the thickening of pancakes, they implement mixed Lagrangian
and Eulerian approaches that are difficult to treat analytically.
Conversely, constructing a realization of the density field requires
to follow particle trajectories numerically, as in an N-body
simulations (though without computing forces at each time-step).

\subsection{Truncated Zeldovich approximation and beyond}
\label{section:tza}

\begin{figure} 
\centering{
\begin{tabular}{ccc}
  \shortstack{(\textbf{a}) Nbody \\ \includegraphics[width=0.3\textwidth]{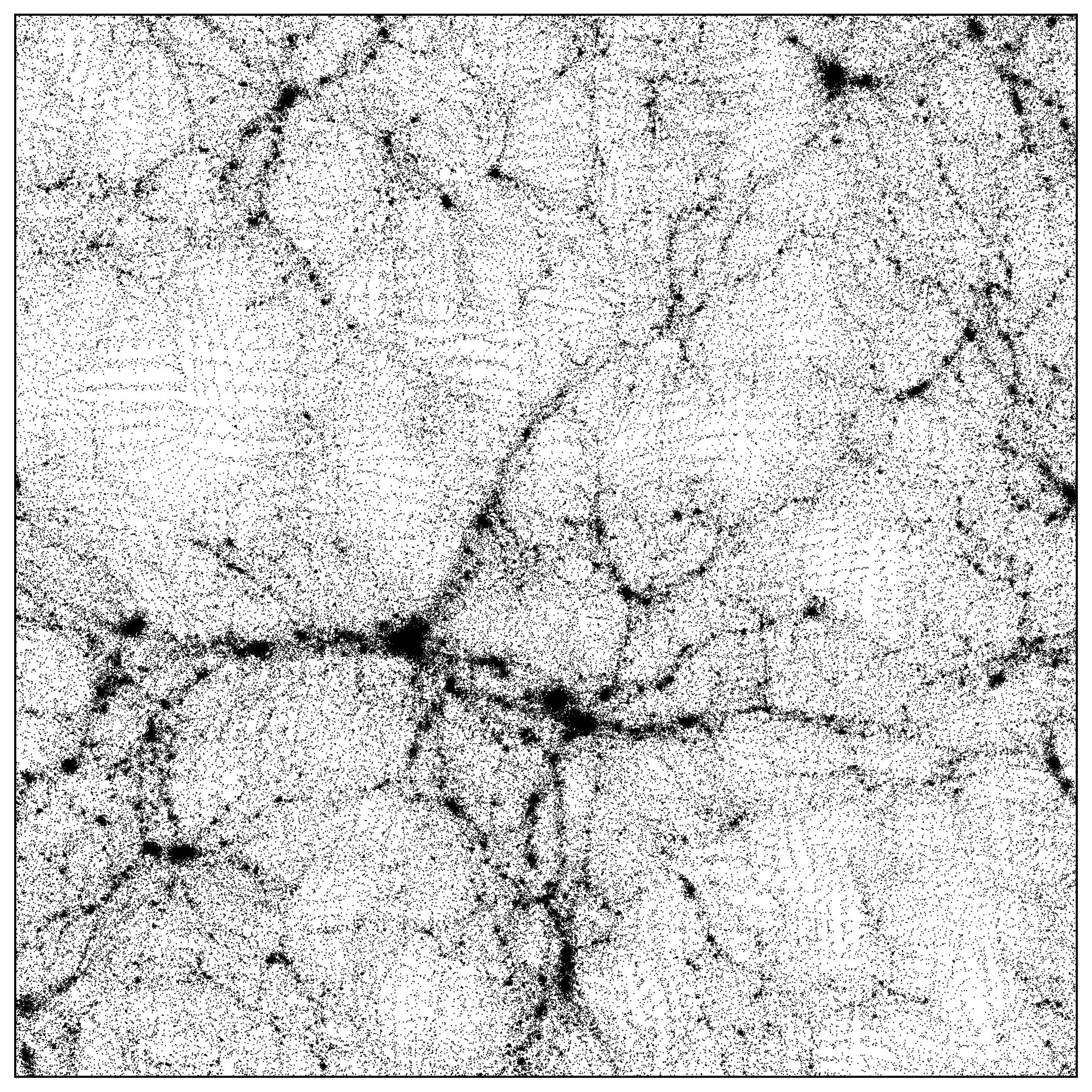}} & 
  \shortstack{(\textbf{b}) TZA \\ \includegraphics[width=0.3\textwidth]{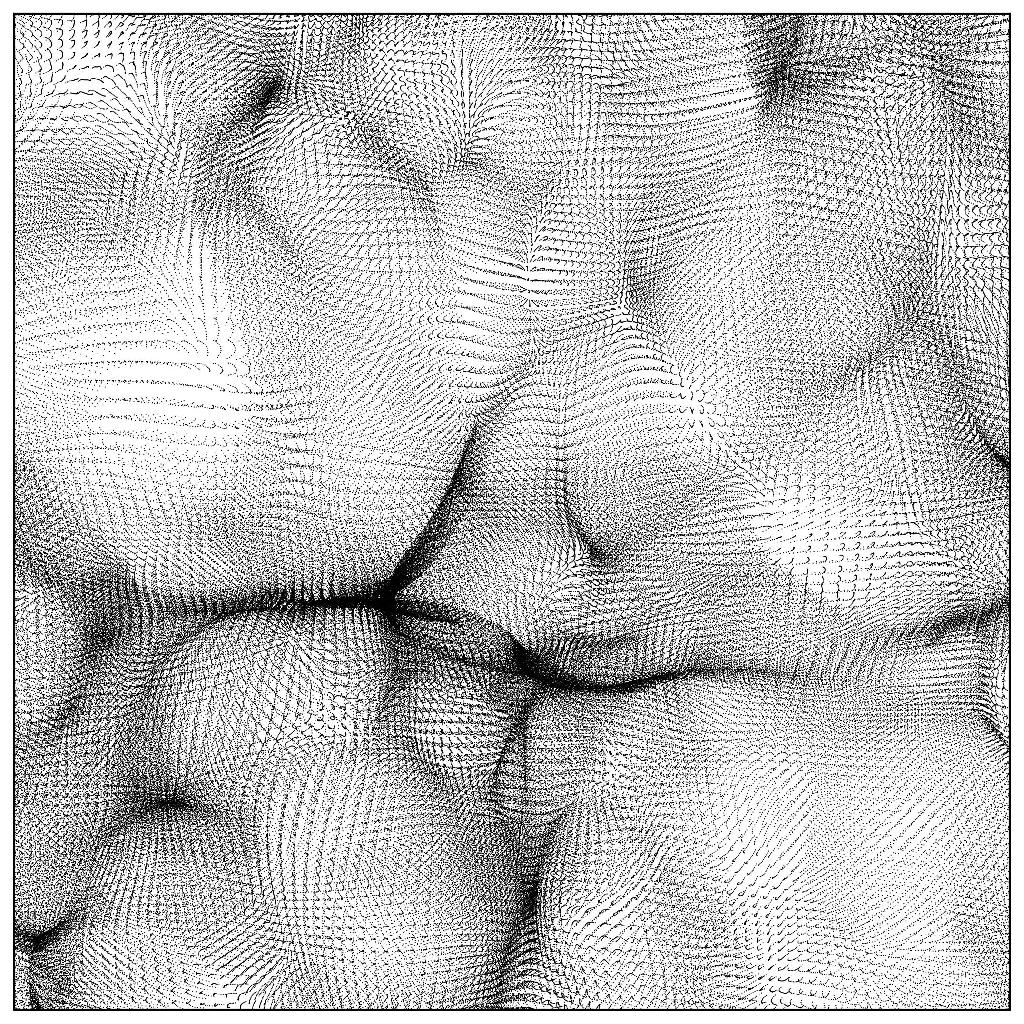}} &
  \shortstack{(\textbf{c}) T2LPT \\ \includegraphics[width=0.3\textwidth]{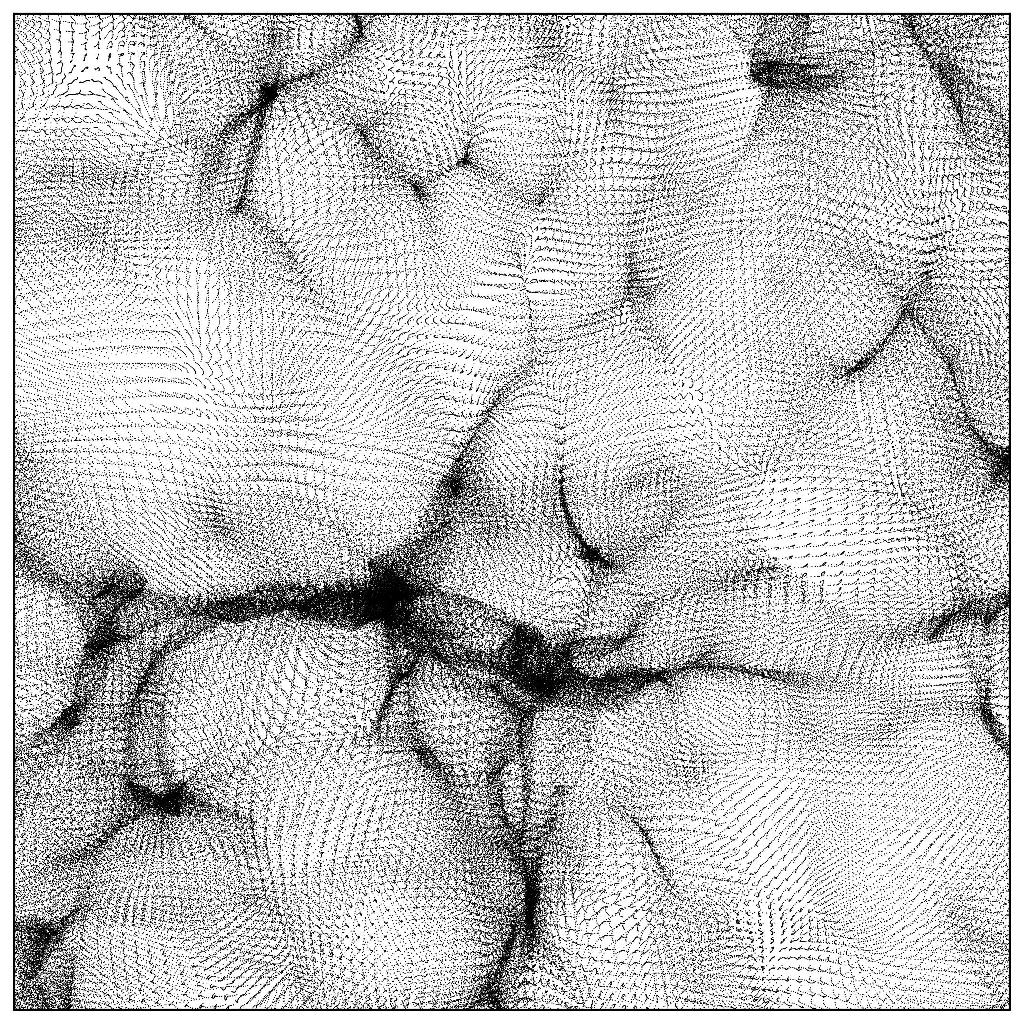}} \\

  \shortstack{(\textbf{d}) ZA \\ \includegraphics[width=0.3\textwidth]{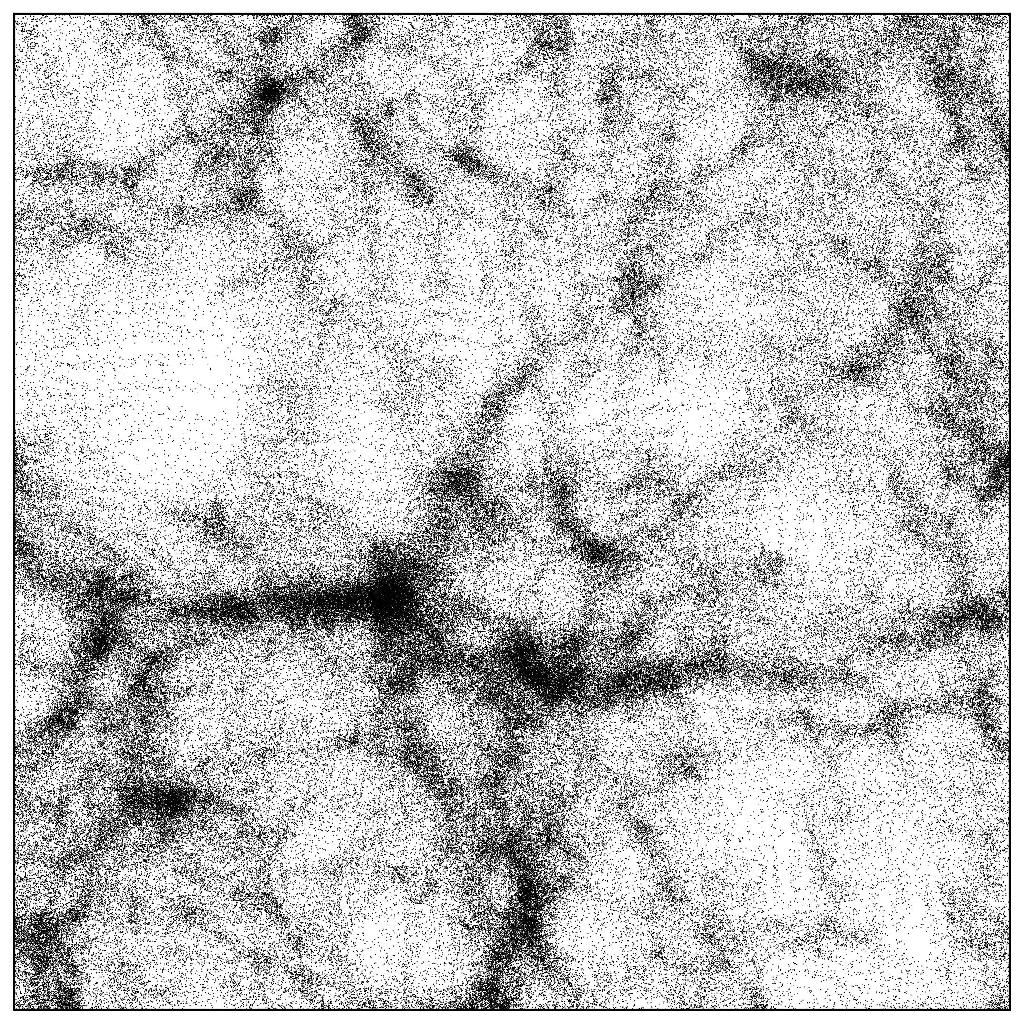}} & 
  \shortstack{(\textbf{e}) 2LPT \\ \includegraphics[width=0.3\textwidth]{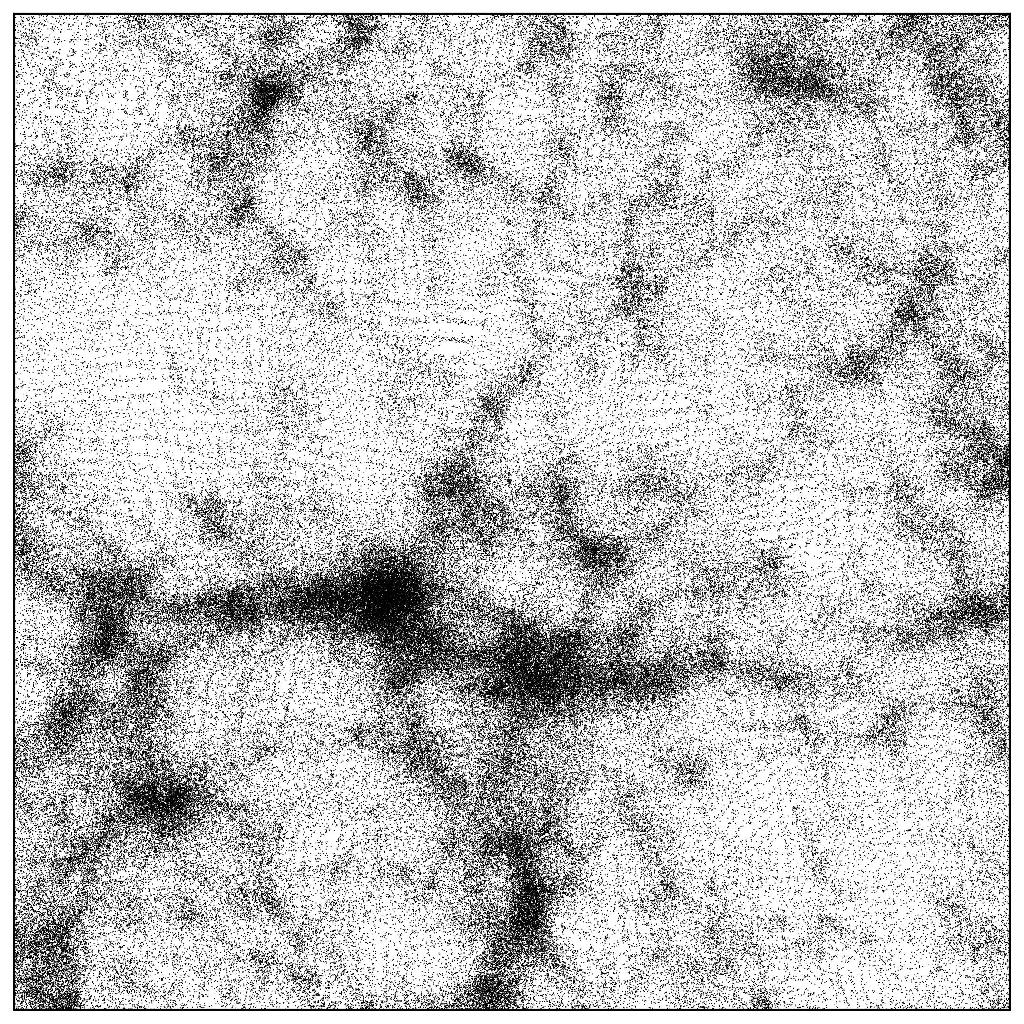}} & 
  \shortstack{(\textbf{f}) 3LPT \\ \includegraphics[width=0.3\textwidth]{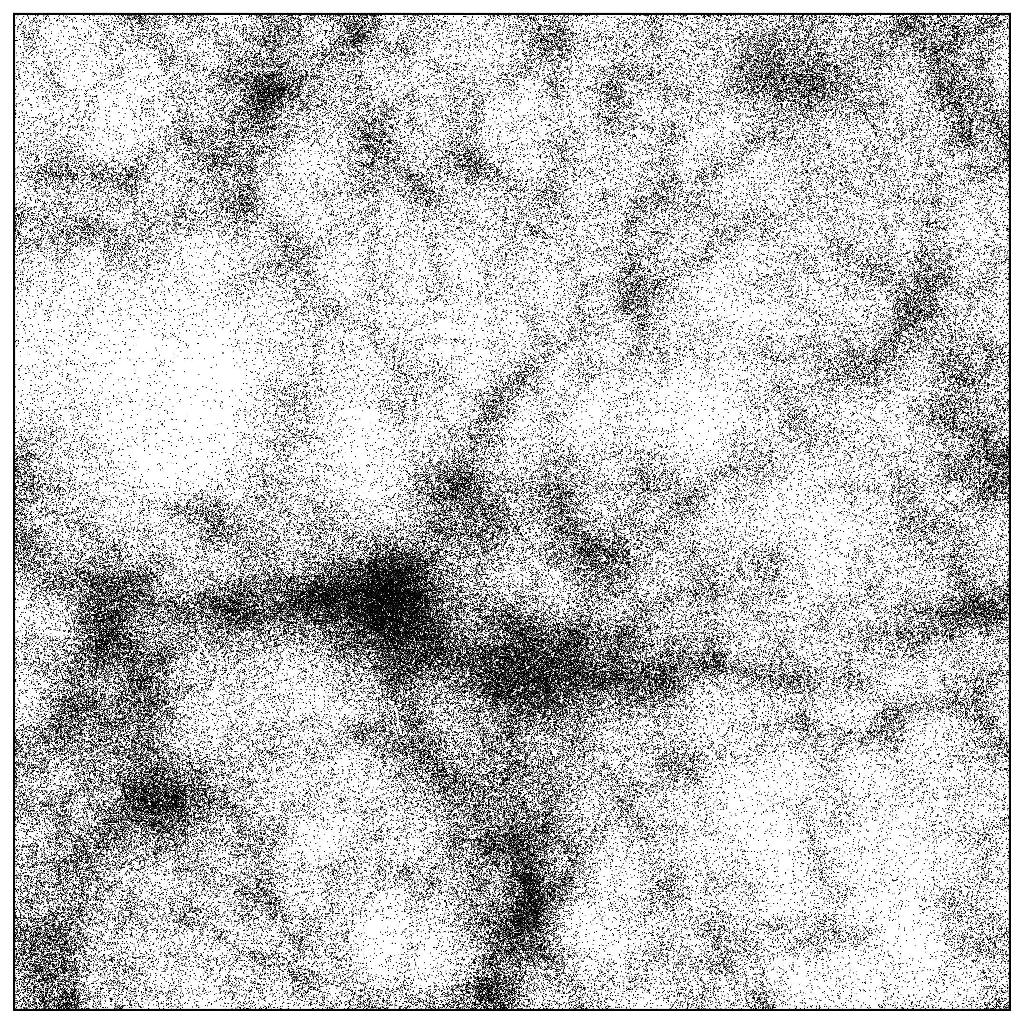}} \\

  \shortstack{(\textbf{g}) COLA \\ \includegraphics[width=0.3\textwidth]{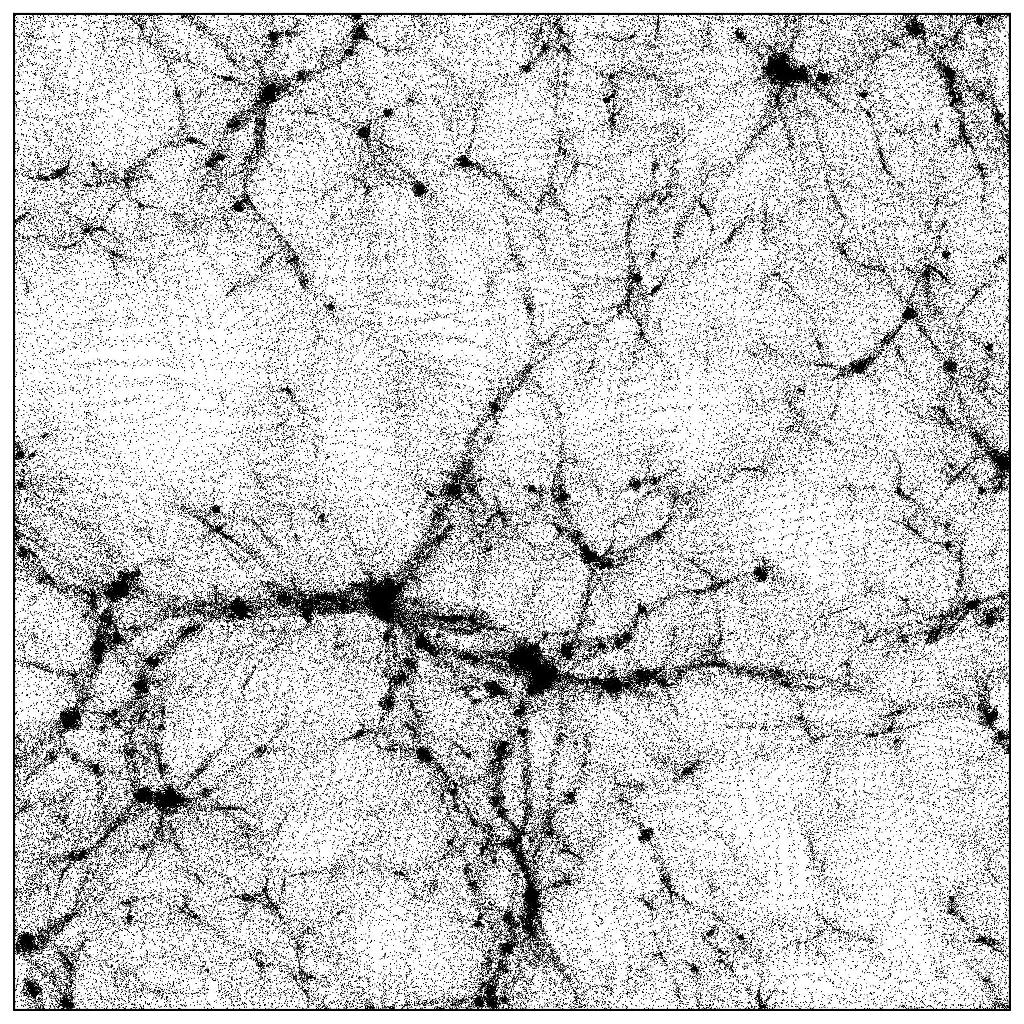}} &
  \shortstack{(\textbf{h}) A2LPT \\ \includegraphics[width=0.3\textwidth]{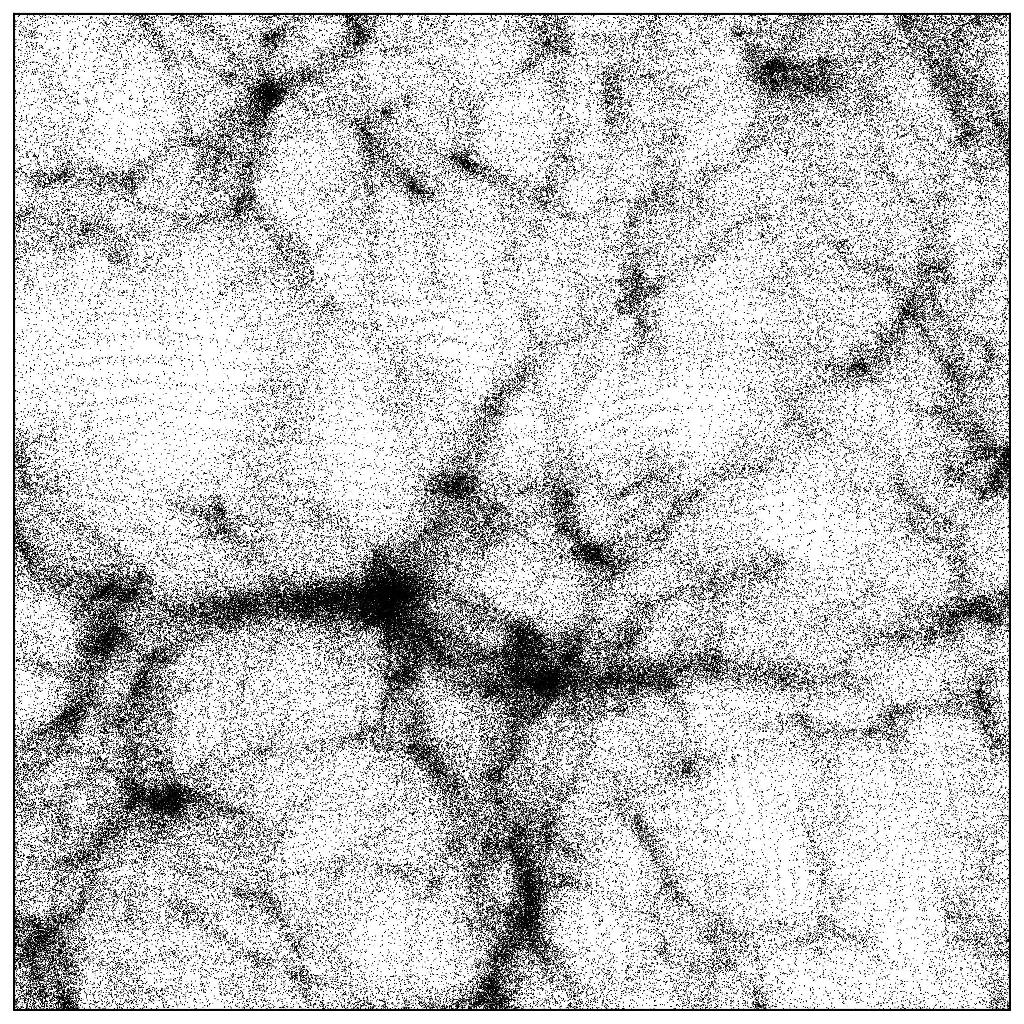}} & 
  \shortstack{(\textbf{i}) A3LPT \\ \includegraphics[width=0.3\textwidth]{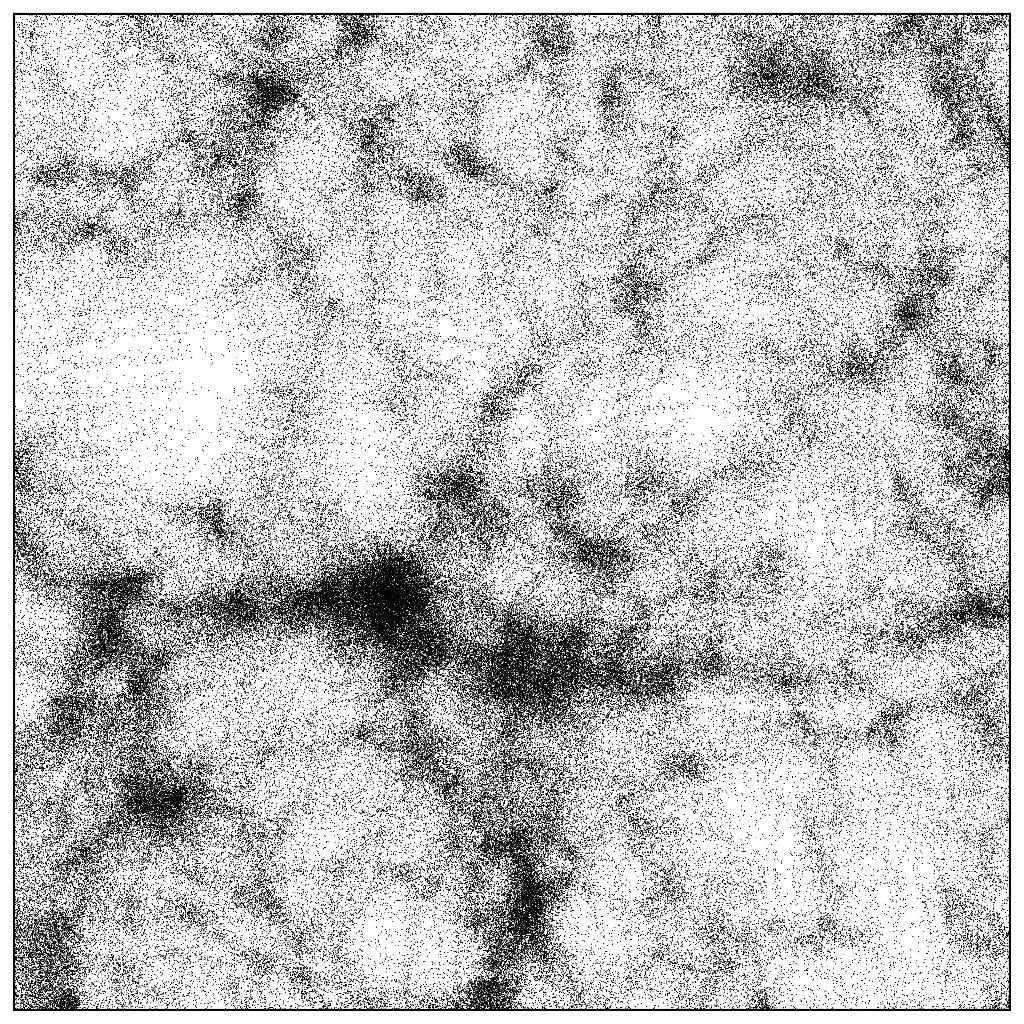}} \\
\end{tabular}
}
\caption{Large-scale structure in a 20 {\Mpch} thick slice of a square
  of 128 {\Mpch}, as recovered by different methods: (\textbf{a})
  N-body simulation; (\textbf{b}) truncated ZA; (\textbf{c}) truncated
  2LPT; (\textbf{d}) ZA; (\textbf{e}) 2LPT;
  (\textbf{f}) 3LPT; (\textbf{g}) COLA; (\textbf{h}) augmented 2LPT;
  (\textbf{i}) augmented 3LPT.}
\label{fig:lss}
\end{figure}

The ZA can be applied to the smoothed density field, and the smoothing
length $R$ can be optimized to obtain the best agreement with
simulations, as proposed by \citet{coles1993}. { The same approach can
be performed with the LPT series at any term (\citet{melott1995}); } this was deepened by
Melott in a series of papers reviewed in \cite{melott1994}. He and his
collaborators compared the predictions of several approximations,
among which truncated ZA, 2LPT and 3LPT, to the results of N-body
simulations by applying them to the same initial density field, and
comparing the smoothed densities on large scales with a standard
cross-correlation coefficient. The effect of smoothing on LPT is to
limit the thickening of structure after orbit crossing, and the best
representation of the N-body density field was obtained for a
smoothing scale $R$ of order of that for which $\sigma^2(R)=1$. This
was done for scale-free initial conditions with much small-scale
power: $P(k)\propto k^n$ and $n\ge-1$. Figure~\ref{fig:lss} shows, in
the second and third panel of the upper row, the large-scale structure
obtained with the truncated ZA (middle panel) and 2LPT (right panel)
approximations. The truncation radius is set at 3 \Mpch. The visual
effect of the truncation is very apparent, but it is difficult to
judge by eye if the truncated versions, that completely miss the
small-scale details and yet show significant thickening of pancakes,
give a better representation of the simulation. Actually they do not;
this can be easily recovered from the original papers of \citet{coles1993} and \citet{melott1995},
where improvements were reported for power-law spectra with growing or
flat slope $n=1$ or 0, while for $n=-2$ the cross correlation
coefficients of straight and truncated ZA were very similar. The slope
of the $P(k)$ used here at $k=2\pi/3$ {\hMpc} is $\sim-2.5$. Today
most people would test the method by checking how this recovers the
power spectrum (or two-point correlation function) of the density
field; we will show this analysis later in
Section~\ref{section:comparisons}.

The truncated ZA method was exploited by \citet{borgani1994} to
generate mock catalogs of galaxy clusters, to be compared with
measurements of the two-point correlation function of galaxy clusters
available at that epoch. A density field was generated with truncated
ZA, and the peaks of the density field were populated with a number of
clusters, so as to reproduce the required number density. The
production of many mock catalogs allowed to estimate sample
variance, while the high speed of the computation allowed to test
several cosmological models. The result was to rule out flat and open
CDM models, while tilted CDM and mixed Cold-Hot DM models were
acceptable; the cosmological constant was not yet fashionable at that
time. This is possibly the first examples of generation of mock
catalogs with an approximate method, used to obtain up-to-date
cosmological information.

\subsection{Reconstruction of initial conditions}
\label{section:reconstruction}

{ Another application of approximate methods that attracted much
  attention in that period is their use to reverse the evolution of
  perturbations, to recover the initial conditions that gave origin to
  a specific, observed large-scale density field. A direct backward
  integration of a density field, known with some significant
  uncertainty, would amplify the noise by treating part of it as a
  decaying mode, but the evolution can be inverted in a much easier
  way by restricting the dynamics to the growing mode. This was
  achieved either by using the Zeldovich approximation (or a higher
  LPT order), as in the time machine of \citet{nusser1992} applied
  to peculiar velocity fields as wall as to the density field traced by
  IRAS galaxies, or the least action principle, as proposed by
  \citet{peebles1989} (see \citet{keselman2016} for a recent
  assessment on this technique). Some review of that line of research
  can be found in the introductions of \citet{monaco1999b}, that
  presented a recursive algorithm to find the Zeldovich or 2LPT
  displacement field that gives rise to an observed density
  distribution, and of \citet{mohayaee2006}. The latter authors
  tested against numerical simulations a Monge-Amp\`ere-Kantorovich
  approach, proposed by \citet{mohayaee2003}, that is a rigorous
  formalization of the two-boundary problem of the inversion of a
  Lagrangian map, subject to a constraint of homogeneity at early
  times and an observational constraint at late times. 

This effort of reversing the growth of structure was paralleled by the
work of \citet{hoffman1991}, who proposed a statistical method to
generate constrained realizations of a Gaussian density field. This
was again conceived as a tool to generate initial conditions for
N-body simulations subject to observational constraints, so as to
reproduce the observed large-scale structure of a galaxy sample,
though no emphasis was given to the aspect of non-linear dynamics. It
is clear that the two dynamical and statistical approaches were to be
joined together at a point.}

\section{The age of precision cosmology}
\label{section:precision}

Measurement of cosmological parameters requires the comparison of
observables, like galaxy clustering in redshift space or galaxy weak
lensing, with theoretical predictions. The next generation of galaxy
surveys
\citep{abell2009,levi2013,laureijs2011,green2012,frieman2013,dawson2016}
will permit to beat down the statistical error on these observables,
thus allowing us to obtain parameters with sub-percent-level
errorbars. Systematics, both in the measurement and in the
theoretical prediction, must be controlled at the same significance
level. Percent accuracy is a challenge even for N-body simulations,
that requires massive tests and close attention to all the numerical
details \cite{heitmann2010,reed2013,schneider2015}, starting from
initial conditions \cite{crocce2006,lhuillier2014,garrison2016}. Approximate
methods, as their name suggests, generally cannot meet such high
requirements of accuracy.

But estimation of confidence levels for parameters requires 
{ (under the simple assumption of Gaussian-distributed errors)}
 to specify
a precision matrix, the inverse of the covariance matrix of the
measurements. This matrix must take into account all known
systematics, including a number of nuisance parameters that must be
marginalized over. A proper estimate of this covariance matrix
typically requires the production of a very large number of mock
catalogs to properly sample the large number of bins that optimally
exploits the available information. Smart techniques
\cite{pope2008,schneider2011,percival2014,paz2015,kalus2016,pearson2016,oconnell2015,padmanabhan2015}
can limit this number, but this is likely to remain as high as several
hundred independent realizations. What is needed here is not high
accuracy in the mean but the correct variance.

As far as galaxy clustering is concerned, accuracy is needed only down
to a scale of a few Mpc. Indeed, DM halos contain a number of galaxies
that increases with halo mass, so while galaxy clustering on large
scales is determined by the clustering of their host halos, small
scales are dominated by the clustering of galaxies that reside in the
same halo \cite{cooray2002}. The point where these 2-halo and 1-halo
terms have equal amplitude is at low redshift at $\sim2-3$ \Mpch, or
at a wavenumber of $k\sim0.5$ \hMpc. Smaller scales are dominated by
the motion of galaxies in the highly non-linear halos, and are
typically not used for cosmology. Few percent accuracy down to a few
{\Mpch} or up to a fraction of {\hMpc} is reachable by approximate
methods.

A feasible strategy to compute the covariance matrix for clustering
statistics is that of producing a few accurate mocks with expensive
N-body simulations, that give the right average value\footnote{ 
  Recently, \citet{angulo2016b} showed that it is possible to use two
  sets of initial conditions, with Fourier mode amplitudes fixed to
  the cosmological mean and opposite random phases, to converge very
  quickly to the average value of several statistics. This can be used
  to speed up convergence on averages.
}, and a large
number of mocks with quick approximate methods, that give a biased
average but allow a much better sampling of the covariance matrix. 
{ One possibility is to straightforwardly put together average from
  N-body and covariance from mocks. A more sophisticated approach is 
given by the
so-called shrinkage technique of \citet{pope2008} (originally proposed in \cite{schafer2005}).} 
The concept is the
following: suppose that we
want to estimate a quantity from a set of $n$ measurements, using two
different models. One model, ${\mathbf u}$, is very accurate and
produces a prediction with negligible bias, but has a high variance
because, for instance, it has many parameters. A second model, the
target ${\mathbf t}$, is simpler and it is subject to a smaller
variance (say, it has fewer parameters), but the simplicity is paid
off by a bias. It is possible to combine the two models to obtain an
optimal estimate:

\begin{equation}
{\mathbf u}^\star = \lambda{\mathbf t} + (1-\lambda){\mathbf u}
\label{eq:shrinkage} \end{equation}

\noindent
Optimal value of the shrinkage parameter $\lambda$ is obtained by
minimizing the so-called risk function, that gives the sum of the
expected mean square errors over all measurements. 
{ This technique can be applied to obtain a good estimate of the
  covariance matrix using as model one or a few full N-body
  simulations, and as target a large number of approximate mocks (as
  done for instance in \citet{delatorre2013}).
  Alternatively, if the structure of the covariance matrix is
  demonstrated to be well represented by approximate methods, one
  could use an analytic model as target and a more
  limited number of approximate simulations as model.}

The need for covariance matrices has triggered a renewed interest for
those methods that are able to produce mock catalogs, relatively
accurate up to a few Mpc scale, in a short time. Below, after
discussing some recently proposed extensions of the foundations
discussed in Section~\ref{section:foundations} and the issue of the
universal mass function, I will review the main methods that have been
used or are being developed for the production of such mock catalogs.

\subsection{Recent development of the foundations}
\label{section:recent}

Development of LPT has not stopped in the '90s, and several new ideas
have recently been proposed. The performance of the first-order LPT
term, the ZA, was recently re-assessed by White \citet{white2014b,white2015}. 
This approximation has been widely used for the
reconstruction of the BAO peak
\citep{eisenstein2007,padmanabhan2009,padmanabhan2012,burden2015}: this is
smoothed by non-linearities in redshift space, that can be corrected
for by using ZA displacements and their implied velocities to track
the redshift space back in time to the Lagrangian space. This same
problem was faced by \citet{mccullagh2012} with a novel second-order
expansion of LPT, based on Fourier-Bessel decomposition. Extensions of
LPT to fourth order were presented by \citet{rampf2012} and
\citet{tatekawa2013}. 
{ \citet{leclercq2013} proposed a remapping technique applied to LPT to recover loss of power at small scales.}
An LPT re-expansion was proposed by
\citet{nadkarnighosh2013}, who extended the approach so as to apply it
to an already evolved configuration. In this case it is possible to
re-iterate it analytically, achieving great accuracy up to orbit
crossing. 
\citet{bartelmann2015}, using a Green's function approach,
proposed an analytic, phenomenological scheme to improve the ability
of ZA to predict particle trajectories. It is based on the evidence
that its validity is limited in time due to orbit crossing, so that
slowing down the growth factor after some time (corresponding to the
time when orbit crossing becomes frequent) produces a better agreement
with simulations. 
Connection of LPT with the statistics of the
large-scale field was treated by many articles, good starting points
to this field can be \citet{tassev2014,sugiyama2014,vlah2015}. An
effective field theory approach to the development of non-linearities
was proposed by \citet{carrasco2012}; the method, initially formulated
for EPT, was extended to LPT by \citet{porto2014} and tested against
simulations by \citet{baldauf2016}. 

An interesting improvement of LPT, named Augmented LPT (ALPT), was
proposed by \citet{kitaura2013}. We have discussed above how higher
orders of LPT increase the accuracy up to orbit crossing, but worsen
the thickening of collapsed structures by giving a non-linear boost to
the drift of particles out of caustics. This is well visible in the
middle row of panels of Figure~\ref{fig:lss}, where the first three
LPT orders are shown. Conversely, relaxation makes particles to
stay bound within a limited region of negligible size with respect to
the large-scale structure. This was formalized by
\citet{bernardeau1994} (see also \cite{neyrinck2013,chan2014}), who noticed that the divergence of the
displacement fields, a dimensionless field called stretching
parameter, should not go below $-3$. \citet{mohayaee2006}  { worked out an approximation}
for the relation of the stretching parameter with density found in
N-body simulations:

\begin{equation}
\nabla_\vecq\cdot \vecS(\vecq) = \left\{
\begin{array}{ll} 3 \left[ \left( 1- \frac{2}{3}D(t) \delta(\vecq)
    \right)^{1/2} -1 \right]
& {\rm if}\ \delta < \frac{3}{2D}\\
-3& {\rm if}\ \delta \ge \frac{3}{2D}\end{array}  \right.
\label{eq:divS}\end{equation}

\noindent
Let's call $\vecS_{\rm SC}$ the solution of this equation for the
displacement field.
\citet{kitaura2013} exploited this property of $\nabla_\vecq\cdot \vecS$
within an approach similar in spirit to the peak-background split. They
defined a Gaussian smoothing kernel $\kappa(\vecq; R)$, with $R\sim4$
\Mpch, to separate large-scale modes, that evolve in a mildly
non-linear way, from small-scale modes, that collapse spherically
following $\vecS_{\rm sc}$:

\begin{equation}
\vecS_{\rm ALPT}(\vecq) = \kappa(\vecq; R) \ast \vecS_{\rm LPT} 
+ [1-\kappa(\vecq, R)] \ast \vecS_{\rm sc}
\label{eq:aug} \end{equation}

\noindent
For the first term 2LPT was used in the original paper, but nothing
prevents it to be applied to 3LPT. Figure~\ref{fig:lss} shows, in two
of the lower panels, the large-scale structure recovered by ALPT at
second and third order. Structures are much neater than the straight
LPT approach, especially in the A2LPT case. Another scheme
based on the stretching parameter, MUlstiscale Spherical ColLapse
Evolution (MUSCLE), was recently proposed by 
\citet{Neyrinck2016}.

The excursion sets approach has been developed beyond the moving
barrier, finally providing merger histories that are approximately
self-consistent and in agreement with simulations
\cite{cole2008,parkinson2008}. A very elegant formulation, based on a
path integral approach, of the excursion set approach in the case of
Gaussian smoothing was proposed by Maggiore and Riotto
\cite{maggiore2010a,maggiore2010b,farahi2013}.

The peak approach has also received further attention. Two extensions
are worth mentioning. Firstly, in the formalism named Confluent System of Peak
Trajectories of \citet{manrique1995,juan2014} peaks in the linear,
Gaussian-smoothed density field are supposed to collapse into halos
following an ellipsoidal collapse. The peak-in-peak problem is solved
by considering nested peaks, i.e. peaks that ``merge'' when the
smoothing radius is changed incrementally. Then, the tree of nested
peaks, obtained by continually varying the smoothing radius, is
transformed into a merger tree for the halos. With some assumptions on
halo virialization, the initial profile of a peak is mapped into the
halo final density profile, recovering the density profile found in
simulations \cite{navarro1996}. Halo density profiles and halo mass
functions are reproduced with a minimal number of free parameters,
showing a very interesting level of predictive power. However, the
construction of the merger tree implies a precise distinction between
halo mergers and accretion of uncollapsed matter, that is not what is
commonly seen in N-body simulations where, at modest resolution,
accretion onto halos is largely merging with unresolved objects.

The other approach that puts together the peak formalism and EPS, the
excursion set peaks approach, was proposed by
\citet{paranjape2012,paranjape2013a}. Using an extension of the
excursion sets approach, proposed by \citet{musso2012}, that can
manage smoothing kernels that are not sharp in $k$-space (and thus
produce correlated trajectories in the $\delta-\sigma$ plane),
these authors were able to formulate an excursion sets approach for
peaks. This allowed them to apply well-known techniques, including the
moving barrier, to obtain the mass function of halos, their
conditional mass function and their bias. This model has been compared
to N-body simulations in \citet{paranjape2013b}, showing that it is
able to reproduce halo bias well beyond the linear bias level.

The correspondence of peaks of the initial density field with real
halos was tested against simulations by
\citet{ludlow2014,borzyszkowski2014}. Looking for scale-dependent
peaks in the initial conditions of an N-body simulation, they tested
many of the basic assumptions on the connection of peaks with halos,
on their ellipticity in the Lagrangian space and on the computation of
their collapse time, and found that while good agreement is achieved
at high masses (with a number of caveats), large-scale tidal fields
deeply impact on the evolution of small halos, especially those
residing within filamentary network, breaking the simple mapping of peaks into
halos.

Extensions of the analytic approach to bias described in
Section~\ref{section:bias} would deserve a long discussion, that is
however beyond the scope of this paper. An idea of the liveliness of
the field, and of the complexities in the definitions of non-linear
terms (in configuration or Fourier space), can be found in the
introduction of \citet{paranjape2013b}, while, in the absence of a
recent review, a general overview can be found in the web page of the
dedicated workshop ``Galaxy Bias: Non-linear, Non-local and
Non-Gaussian''\footnote{\tt http://indico.ictp.it/event/a12214/}. Finally,
a general relativistic formulation of the problem of bias can be found
in \citet{baldauf2011}.

It is useful to stress that, while halo bias can be in principle
accurately predicted by N-body simulations, galaxy bias relies, even
on large scales, on the ability to predict how a given selection of
galaxies populates DM halos. In other words, it is unrealistic to
think of predicting the level of bias of a given galaxy population
with percent accuracy. Fortunately, this does not hamper the
possibility to constrain cosmological parameters. One reason for this is that
the main constraint is the position of the BAO peak, whose accurate determination does
not depend on the normalization of power spectrum or 2-point
correlation function. Moreover, galaxy bias can be measured directly
from the data using higher-order correlations
\cite{McDonald2006,gilmarin2015}.

{ The reconstruction of the initial conditions that reproduce a
  given observed region of the Universe and the constrained
  realization approach of \citet{hoffman1991}, discussed in
  Section~\ref{section:reconstruction}, have ultimately evolved into a
  Bayesian inference approach (see \citet{kitaura2008}). As a
  significant difference with previous approaches, this method adopts
  a forward modeling of the evolution of perturbations, thus avoiding
  the problems connected to the inversion of the evolution of
  perturbations. This line of research was pursued by several groups,
  proposing the methods KIGEN \cite{kitaura2013b,kitaura2012}, ELUCID
  \cite{wang2014}, BORG \cite{jasche2013,jasche2015} and the the Constrained
  Local UniversE Simulations (CLUES) project \cite{gottlober2010}. The
  program is ambitious: Bayesian techniques are used to optimally
  sample the huge parameter space of initial conditions, evolving them
  to the final configuration state in order to
  recover the most likely initial density field and its confidence
  level. 
  The technique is analogous to that proposed for the Cosmic Microwave Background
  \cite{wandelt2004}, with the relevant difference that the density
  field is 3D and must be non-linearly evolved; the non-linear
  evolution is exactly where approximate methods come into play. }

\subsection{The universal mass function}
\label{section:universal}

A consequence of the PS approach is that the halo mass function can be
recast into a ``universal'' function that does not depend on time or
cosmology. Let's define $\nu=\delta_c/\sigma$; at any time and for any
cosmology, mass elements with $\nu>1$ are expected to collapse, so
that this quantity can be seen as the ``peak height'' of collapsing
objects, though no peak condition is set here. Taking the derivative
of both terms of equation~\ref{eq:PS2} with respect to $\nu$, and
including the fudge factor of two, we obtain:

\begin{equation}
\frac{1}{\bar\rho}\, M n(M) \times \frac{1}{\nu}
\frac{dM}{d\ln\nu} = \sqrt{\frac{2}{\pi}}
\exp\left(-\frac{\nu^2}{2}\right)
\label{eq:univPS}
\end{equation}

\noindent
The right hand side is a function only of the peak height $\nu$, so
the left-hand term cannot depend on time or cosmology. The mass
function can thus be expressed as:

\begin{equation}
\nu f(\nu) = \frac{1}{\bar\rho}\, M n(M) \frac{dM}{d\ln\nu}
\label{eq:univMF}
\end{equation}

In the following we will define halos using either the percolative
Friends-of-Friends (FoF) algorithm, based on linking particles below a
certain distance, typically taken to be 0.2 times the average
inter-particle distance, or the Spherical Overdensity (SO)
algorithm, where halos are defined as spheres around density peaks,
with radius such that the average (over-)density of the halo is equal
to 200 times the average or the critical density.

In the case of a power-law power spectrum in an Einstein-de Sitter
universe, it must be possible to express the halo mass function as a
universal function of $\nu$. In other cases, this expectation is not
guaranteed to be true. Moreover, different halo definitions will give
different evolutions with time, so that the resulting mass functions
cannot all be universal. This was addressed by 
\citet{despali2015}, who found that universality of the mass function
of simulated DM halos is obtained provided that the halo virial radius
and the critical collapse density $\delta_c$, contained in the
definition of $\nu$, are scaled with redshift according to the
prediction of spherical collapse.

If the mass function is universal, then one can use simulations to
find the $f(\nu)$ function to any required accuracy: the mapping from
a mass range to a $\nu$ range depends on time, so a good coverage of
$\nu$ can be obtained simply by considering halos at different times.
A suite of simulations at several mass resolutions will thus easily
over-sample the $f(\nu)$ curve and give the correct numerical
solution; to be more precise, one solution for each halo definition.
Starting from the seminal 1988 paper of 
\citet{efstathiou1988}, where the ``multiplicity function'' (closely
related to $f(\nu)$) of simulations was shown to be relatively well represented
by the PS formula, many simulation programs have been carried out to
quantify the mass function of halos to within a good accuracy, fitting
it with universal functions
\citet{sheth1999,jenkins2001,warren2006,reed2007,angulo2012} or
claiming a break of universality \citet{tinker2008,crocce2010,manera2010,bhattacharya2011,kim2015}. In
\citet{courtin2011} the mass function is claimed to be universal, but
at a level below the accuracy of their N-body simulation, so they
provide a universal fit to they results. In \citet{watson2013} the
mass function is found to be almost universal for FoF halos, but
significantly non-universal for SO halos \citep[as in][]{tinker2008}.

\begin{figure}
\centering{
\includegraphics[width=15cm]{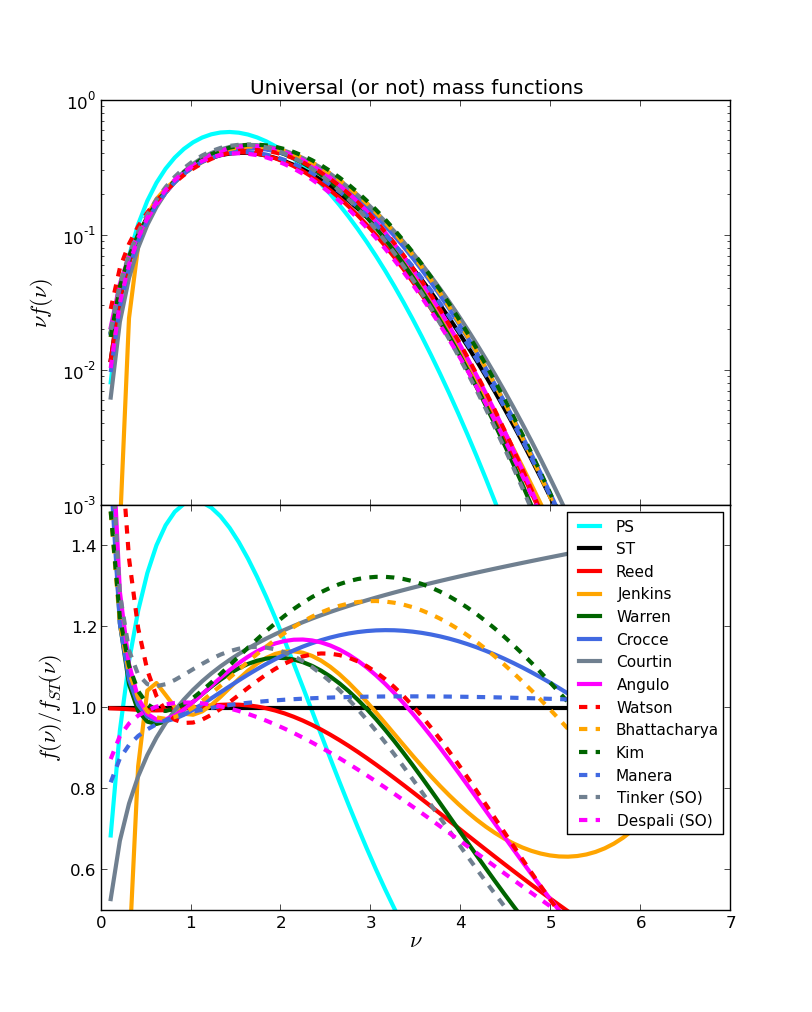}
}
\caption{ ``Universal'' representation of the mass function $f(\nu)$:
  the figure reports various fits of simulations proposed in the
  literature and cited in the text, at $z=0$. The lower panel gives residuals with respect to the
  Sheth and Tormen (ST) mass function.}
\label{fig:universalMF}
\end{figure}

Figure~\ref{fig:universalMF} shows several fits of the mass functions $f(\nu)$
of simulated DM halos, proposed by the papers cited above, at $z=0$. The lower panel gives the
residuals with respect to the \citet{sheth1999} (ST) one, taken as a
reference. Besides PS, the other functions are fit on the numerical
mass function of FoF halos, with the exception of \citet{tinker2008},
that is fit on SO halos; we report it to illustrate the difference
between the two halo finders. Also, \citet{reed2007} is calibrated on
relatively small halos at very high redshift, $z\sim10$, while other
fits are calibrated on more massive halos at lower $z$. While keeping
in mind that any value at $\nu\ge4$ is an extrapolation, it is clear
that agreement at $\sim5$ \% level between the most recent determinations of FoF halo mass functions is achieved only at $\nu\sim1-2.5$.

\subsection{Lagrangian methods to produce mock catalogs}

We can broadly divide the approximate methods for the generation of
mock catalogs in two classes. In the first class the evolution of a
linear density field, sampled on a grid, is followed with the intent of
understanding, deterministically, which particles gather into DM
halos. I will call them ``Lagrangian'' methods. In the second class a
non-linear density field is produced, then a bias model is used to
stochastically populate the density field with halos. I will call them
``Bias-based'' methods.

I start discussing the Lagrangian methods. In fact, some of these are
just quick N-body solvers, where accuracy is compromised to achieve
maximal speed.

\subsubsection{PINOCCHIO}

The PINpointing Orbit Crossing Collapsed HIerarchical Objects
(PINOCCHIO) was developed by Theuns, Taffoni and myself in 2002
\citep{Monaco2002a,Monaco2002b,Taffoni2002}. This scheme puts together
three of the foundational theories discussed above: ellipsoidal
collapse is used to estimate the time at which a particle is expected
to collapse, in the orbit crossing sense; LPT is used to approximate
the evolution of a homogeneous ellipsoid, with a correction to
properly reproduce spherical collapse, and to displace halos from
their Lagrangian position; excursion sets theory is used to deal with
multiple smoothing radii.

The algorithm works as follows. The linear field is Gaussian-smoothed
on several scales, to sample the trajectories (of inverse collapse
times versus $\sigma(R)$) and thus determine the earliest collapse
time for each particle (grid cell). These are then sorted by collapse
time in chronological order, and grouped into halos using an algorithm
that mimics hierarchical clustering. This includes a treatment of
filaments, made up by particles that have collapsed in the orbit
crossing sense, but do not belong to any halo. Halos grow by accretion
of single particles and by merging with smaller halos.

The tree of merging events is recorded during the halo construction
algorithm and output at the end, with continuous time sampling and
without any need to write full outputs at tens of different redshifts.
These are very useful for galaxy formation purposes; the MORGANA SAM
of galaxy formation developed by Fontanot and myself
\citep{Monaco2007} is based on merger trees produced by PINOCCHIO.
These are notably ``well-behaved'', in the sense that events like
halos that merge and then split again, halo mass never decreases with
time, and backsplash substructures getting in and out a halo are
absent.

It is also possible to compute, during the halo construction
algorithm, the time at which a halo passes through a past light cone,
centered on some position within the box. This extension has been
recently developed, taking into account the periodic replications of
the box to tile the volume needed to contain the light cone. The
catalog is directly output at the end, again with no need of writing
several complete outputs.

Recent development of the code has been presented by
\citet{monaco2013} and \citet{Munari2016}. The code has been
demonstrated, since 2002, to be able to reproduce mass and accretion
histories of halos (see references in \citet{monaco2013}) with a good
accuracy, even at the object-by-object level. Halo clustering relies
on the ability of LPT to place halos in their final position; with
3LPT accuracy in the power spectrum to within 10\% can be achieved at
$k\sim0.5$ in redshift space, especially at a relatively high redshift
($z\sim1$). The code, in the 2013 version, has been used to create
mock catalogs for the VIPERS survey \cite{delatorre2013} { and is being used for the Euclid preparatory science.}

\subsubsection{PTHALOS}

The original Perturbation Theory Halos (PTHalos) scheme, proposed by
\citet{scoccimarro2002} in 2002, is based on the idea of producing a
density field on a grid with 2LPT, then compute densities on a coarser
grid and use an algorithm based on the conditional mass function
(Section~\ref{section:bias}) to populate each small volume with halos.
In this case, extreme minimization of computing time is achieved.
Because the application of the conditional mass function is in fact a
bias scheme, this method should stay in the bias-based class. However,
the more recent implementation of PTHalos, developed by 
\citet{manera2013}, is based on a different approach. After creating
the 2LPT density field, a FoF algorithm is run on the particle
configuration, with a linking length that is suitably increased to
allow for the lower average density of halos due to the thickening of
pancakes. This linking length is estimated using an algorithm based on
spherical collapse. Halo masses are then rescaled to force them to
follow a specified mass function. Despite predictiveness on the mass
of halos is lost, this code is predictive on halo bias and clustering.

PTHalos has been thoroughly used for producing mock catalogs and
covariance matrices for the BOSS survey
\cite{dawson2013,manera2013,ross2012,manera2015}, while the latest
products in that collaboration were done using more advanced methods
that will be described below.

\subsubsection{Methods based on the particle-mesh scheme}

One reason why methods like PINOCCHIO or PTHalos are able to produce a
good approximation of a simulation in a much shorter time is due to
the fact that they do not attempt to resolve the inner structure of
halos, where an N-body would spend most time in integrating the very
non-linear orbits of halo particles. A great speed-up can be achieved
also in N-body simulations by compromising on resolution. A
straightforward way to achieve this is to use the classic
particle-mesh (PM) scheme with few time-steps. 
{ As a matter of fact, a code implementing the PM scheme is an
  N-body code. However, in this text I will reserve the term
  ``N-body'' to those codes that give a detailed representation of
  forces at the particle-particle level, down to a softening scale.
  The PM scheme does not achieve this level of detail, and, in fact,
  the community is using it mainly as an approximate method. 
This scheme,
illustrated in textbooks like \citet{mo2010} or \citet{hockney1981},
is based on the idea of solving the Poisson equation for the
gravitational potential on a grid using Fast Fourier Transforms
(FFTs), thus changing an $N^2$ problem into a much more manageable
$N\log N$ one. { Once the Fourier transform of the potential has
  been computed, three backward FFTs (or one FFT and a finite difference scheme)
  will then give the force on each grid point, and the motion of
  particles will be updated with a leapfrog
  scheme.}

The disadvantage of the PM integration scheme is that its resolution
is set by the grid size, so to resolve small scales a very fine grid
is needed, making memory requirements very high. In fact, high
resolution is needed only in the small part of the volume occupied by
high density peaks, so the scheme is far from optimal. But if the
large-scale density field is needed, then a PM code with a coarse
grid, of the order of the inter-particle distance, will produce in few
tens of time-steps a very good approximation of the results of a full
N-body simulation.

PM schemes, being N-body solvers, generate a distribution of
particles. DM halos must be extracted using a halo-finder code,
typically the FoF one (possibly with some rescaling of the linking length to
correct for the lack of resolution at small scales). The overhead due
to FoF is typically reported to be modest. Notably, the recovery of DM
halos puts strong constraints on the mesh used to compute the density.
A mesh as large as the inter-particle distance, like the one used for
the LPT-based methods mentioned above, is typically too coarse to
recover the small halos, $1/2$ or better $1/3$ of the inter-particle
distance is sufficient to recover halos with hundreds of particles
\cite{koda2016,izard2016,feng2016}. This way, memory requirements
grow, as the number of mesh cells, by a factor of 8 or 27 (and CPU
requirements by a larger factor) with respect to the original grid.

PM-based methods were proposed by \citet{merz2005} (PMFAST) and
\citet{white2010} (Quick Particle Mesh, QPM, used in \cite{white2014}). The QPM code
was used by \citet{white2015b} to generate mock catalogs for the BOSS
survey \cite{beutler2014}. A PM code was used in the ELUCID code
mentioned in Section~\ref{section:recent} for the fast evolution of a
set of initial conditions \cite{wang2014}. A lower limit on the number
of time-steps needed to achieve numerical convergence is given by the
requirement of recovering the linear growth of structure. Accurate
results, at per-cent level on large scales, are obtained with tens of
time-steps \cite{feng2016,izard2016}. Two techniques have been
proposed to force particle trajectories to recover the linear growth
in the case of few time-steps, thus improving the numerical
convergence of the code. The first one was devised by
\citet{Tassev2013}, who exploited the fact that 2LPT is already a good
description of the evolution of large-scale structure. They proposed
the COmoving Lagrangian Acceleration (COLA) algorithm, where the
equations of motion are solved in a gauge that is comoving with the
2LPT solution, computed on the initial conditions as usual.
Very good convergence of this modified PM scheme is obtained with $\sim10$
time-steps, so the time overhead of computing 2LPT is more than compensated
by the quicker convergence. However, the solution of 2LPT impacts on
memory requirements, so if a mesh of $1/3$ of the
inter-particle distance is used in order to recover small halos, 
memory is the main
limitation factor of this method.
COLA has been parallelized by several groups
\cite{howlett2015a,koda2016,izard2016}, and used in the production of
mock catalogs of the Sloan Digital Sky Survey main galaxy sample
\cite{howlett2015b}, of the WiggleZ survey
\cite{drinkwater2010,kazin2014} { and in Bayesian inference of the SDSS \cite{leclercq2015}}. Finally, an extension of the COLA
algorithm to standard N-body codes, sCOLA, has recently been proposed
\cite{tassev2015}.

The second technique to recover the correct linear growth of structure
has been proposed by \citet{feng2016} with their FastPM scheme. They
pointed out that if the time-step is large, the assumption of constant
velocity during the time-step becomes inaccurate. They re-defined the
kick and drift operators of the leapfrog scheme assuming that the velocity evolves following
the Zeldovich approximation, thus forcing the linear growth to be
recovered even with few time-steps. They showed that even 5 time-steps
are sufficient to improve significantly with respect to the 2PLT
solution. }


\subsection{Bias-based methods}

Methods belonging to this class generate a large-scale density field
with one of the variants of LPT discussed above, then implement a
sophisticated bias model to populate the density field with halos,
calibrated on some big simulation.

\subsubsection{PATCHY}

The matter density field $\rho_{\rm M}(\vecx)$ obtained with ALPT
(Section~\ref{section:recent}) is used by the PATCHY (PerturbAtion
Theory Catalog generator of Halo and galaxY distributions) algorithm,
presented by \citet{kitaura2014}, to create a distribution of halos
that fit a given simulation at the 3-point level. This is done first
by constructing a phenomenological, stochastic bias model. The halo
density field $\rho_h$ is modeled as a power-law of the matter density
$\rho_{\rm M}$, plus a threshold and an exponential cutoff:

\begin{equation}
\rho_h = f_g \theta(\rho_{\rm M}-\rho_{\rm th}) \exp\left[-\left(\frac{\rho_{\rm M}}{\rho_\epsilon}\right)^\epsilon
\right] \rho_{\rm M}^\alpha(\rho_{\rm M}-\rho_{\rm th})^\tau
\end{equation}

\noindent
Here $\rho_{\rm th}$, $\alpha$, $\epsilon$, $\rho_ \epsilon$ and
$\tau$ are parameters, while the normalization $f_g$ is obtained by
fixing the total number of objects in the sample. This halo density
field is then sampled with a dispersion larger than Poisson to
reproduce stochasticity of bias, possibly induced by non-local terms.
This introduces a further parameter. Peculiar velocities are computed
as the sum of ALPT contribution plus a Gaussian dispersion that is
scaled with matter density, introducing two more parameters. The
assignment of halo masses to halos, so as to correctly reproduce the
halo-dependent bias, is performed with a dedicated Halo mAss
Distribution ReconstructiON (HADRON) code \cite{zhao2015}; the code is
also able to assign stellar masses to the galaxies with which the halo
has been populated. All these parameters are calibrated on a large
simulation with sufficient statistics of halos. The requirement is to
match the power spectrum and the univariate halo probability
distribution function. The resulting catalog provides a good match of
halo clustering up to three-point statistics \cite{kitaura2015}.

This code has been used to produce a very large number, 12288, of mock
catalogs of the BOSS survey, and this set is currently used to obtain
covariance matrices and perform parameter estimation
\cite{kitaura2016}.

\subsubsection{EZmocks}

The Effective Zeldovich approximation mock catalog (EZmocks) code,
by \citet{chuang2015}, takes a step back with respect to PTHalos and
PATCHY by starting from the density field as predicted by the simplest
ZA, and relies on a bias model to best reproduce clustering at the
three-point level. Deterministic bias is recovered by rescaling the
PDF of density, obtained with ZA, to match that of halo density
obtained with a big simulation. Then densities are subject to a
scatter so as to broaden the PDF as follows:

\begin{equation}
\rho_s(\vecx) = \left\{
\begin{array}{lll}
\rho_0(\vecx) [1+G(\lambda)] & {\rm if} & G(\lambda)\ge0\\
\rho_0(\vecx) \exp[{G(\lambda)}]  & {\rm if} & G(\lambda)<0
\end{array} \right.
\label{eq:ezmocks1}
\end{equation}

\noindent
where $G(\lambda)$ is a Gaussian-distributed random number with
standard deviation $\lambda$. This density field is further subject to
a threshold $\rho^{\rm low}$, below which it is set to zero, and a
threshold $\rho^{\rm high}$, above which it is set constant. The
obtained field is then Poisson-sampled with halos, and their power
spectrum is fixed by enhancing small-scale power by a factor $(1+Ak)$
and boosting BAOs by a factor of $\exp(k/k_*)$. As in PATCHY,
velocities are computed as the sum of a large-scale term and a
Gaussian dispersion. All these parameters are fit by requiring a good
reproduction of halo power spectrum and bispectrum. Halo masses, and
galaxy stellar masses once halos are populated, can be again obtained
by the HADRON code mentioned above; as a matter of fact, the same
group is developing the three codes PATCHY, EZmocks and HADRON, and
this last one applies as well to the outputs of the two mock
generators.

\subsubsection{Halogen}

Another approach, named Halogen has recently been devised by
\citet{avila2015}. In this case a 2LPT density field is generated on a
grid and resampled on twice the inter-particle distance, while a set
of halos is produced that samples a given analytic mass function but
are not immediately assigned to a given position. A cell is then
chosen with a probability that is a power law of its mass density,
$P\propto \rho_{\rm M}^\alpha$, and a halo is placed on a random 2LPT
particle within that cell, provided it does not overlap with a
previously chosen one, in which case the choice of the cell is
repeated. The cell mass is decreased by the halo mass, to force mass
conservation. Halo velocities are then assigned, boosting the velocity
dispersion within a cell to reproduce that measured in a given
reference simulation. This procedure is repeated a few times,
addressing each time halos in a specified mass bin and using a
different $\alpha$ parameter for each bin. These parameters are
calibrated by requiring a good fit of halo power spectrum.

\section{Comparison of methods}
\label{section:comparisons}

The two sets of methods described above, the Lagrangian and bias-based
ones, have different
characteristics. Lagrangian methods attempt to forecast the formation
of halos at the object-by-object level, using either (for PINOCCHIO)
ellipsoidal collapse and an algorithm that can be considered as a halo
finder in a loose sense, or a standard halo finder like FoF in the
other cases. These codes are very predictive, because they attempt to
find simulated halos at the object-by-object level (with PTHalos being
less performing from this point of view) and can be calibrated in a
cosmology-independent way. { In fact, PM-based codes including COLA are
quick N-body solvers, where 2LPT (or a redefinition of kick and drift operators)}
greatly helps to
achieve convergence in few time steps. However, resolving small halos
requires accuracy below the inter-particle distance, so FFTs must be
run on $1/2$ or $1/3$ of the inter-particle distance, and this greatly
impacts on memory requirements and limits the feasibility of massive
runs. Within this class, PINOCCHIO gives advantages like merger
histories and past light cone with continuous time sampling and no need to run FFTs below
the interparticle distance, but its accuracy in placing halos is
limited by LPT. Very reasonably, the higher accuracy of PM-based
methods is payed off by higher computational cost.

Bias-based methods, conversely, are far quicker. The need of smoothing
is implemented by using a grid of order $\sim2$ \Mpch, so the mass
resolution of the needed density field is modest and very large boxes
can be easily run. Smaller halos can be easily produced, as long as
their bias is known, just by producing a larger number of objects, the
bottleneck lying in the production of the reference simulation and the
calibration of the parameters that require many evaluations of the
clustering statistics. As a consequence, they can be used to generate
a very large number of realizations of the same cosmology in a very
short time. The negative side is that they have less predictive power,
because for each cosmology and sample definition they need to be
calibrated on a large simulation, and they do not predict the halo
mass function or halo merger histories.

\begin{table}[H]
\caption{Estimated cost of one realization of 4 $h^{-1}$ Gpc box where
  halos of $10^{12}$ M$_\odot$ are resolved. {The last column gives the
  estimated cost of running 1000 realizations.}}\label{table:costs}
\small \centering
\begin{tabular}{llllll}
\toprule
\textbf{Method} & \textbf{Memory} & \textbf{N. cores} & \textbf{elapsed time} & \textbf{CPU time} & \textbf{1000 realizations}\\
\midrule
EZmocks & 40 GB & 16 & 7.5 m & 2 h { + calibration N-body} & 822,000 h\\ 
PATCHY & 40 GB & 16 & 7.5 m & 2 h  { + calibration N-body} & 822,000 h\\ 
PINOCCHIO & 14 TB & 2048 & 30 m & 1,024 h & 1,024,000 h\\
COLA & 33 TB & 4096 & 2.5 h & 10,240 h & 10,240,000 h\\
{ N-body (HugeMDPL)} & 6.5 TB & 2000 & 410 h & 820,000 h & 820,000,000 h \\
\bottomrule
\end{tabular}
\end{table}

Within the Euclid collaboration, I have tried to estimate the time
needed to run a single mock for some of these methods. Clearly, this
exercise is very inaccurate because a proper scaling test is not
performed on a single machine, but it can catch the order of magnitude
of the needed resources. The assumed configuration is a 4 $h^{-1}$ Gpc
box sampled with $4096^3$ particles, where $6\times10^{11}$ $h^{-1}$
M$_\odot$ halos are resolved with only 10 particles.
Table~\ref{table:costs} gives the estimated costs for four codes,
namely EZmocks, PATCHY, PINOCCHIO and COLA (with a grid size of $1/3$
the inter-particle distance). 
Here we give the total memory required,
the number of cores (assuming 8 GB of memory per core), the elapsed
time and the CPU time. 
{ These are compared to the cost of a simulation that samples the same volume with the same number of particles. As an example, I chose
the HugeMDPL simulation of the MultiDark suite \cite{klypin2016}; the data (kindly communicated to me by Gustavo Yepes) refer to the simulation run on the SUPERMUC facility with a highly optimized version of GADGET; let's bear in mind that the other estimates are relatively conservative. From this table, we see that PINOCCHIO is quicker by a factor of $\sim1000$, COLA by a factor of $\sim100$.

In the table I stress that to the cost of the first two methods 
one should add the cost of calibration on a massive simulation. 
Anyway, the
increase of the cost of one run going from the fastest to the most expensive
methods is striking, and gives a good idea of the cost of increased
accuracy and predictive power. 

The last column gives the cost of running 1000 realizations. In this
case I added to the bias-based methods the cost of running a
simulation like he HugeMDPL (but not the runs needed to calibrate the parameters). Costs range from 1 to 10 million hours,
reachable with standard grants of cpu time, while the cost of 1,000
full N-body simulations is clearly huge.}

\subsection{The nIFTy comparison project}

A comparison of different methods was performed in 2014, within the
nIFTy comparisons program of Knebe and Pearce, led by Prada and Chuang
\citep{nifty}, to which many of the people involved in these codes
participated (including myself). A reference catalog was provided
using BigMultiDark simulation \cite{klypin2016} of a 2.5 Gpc/h box
sampled with $3840^3$ particles. Halos more massive than
$\sim10^{13}$ {\Msunh} were selected, for a number density of $3.5\times
10^{-4}\ h^3\, {\rm Mpc}^{-3}$, consistent with the BOSS CMASS sample
\cite{reid2016}. Two different halo finders were considered (FoF and
SO), so the exact mass threshold depends on this choice. A white noise
of the linear density field, re-sampled on $1920^3$ particles, was
distributed to the participants in order to reproduce the same
large-scale structure (though some groups did not use it for technical
issues). Predictions of clustering were produced with PTHalos,
PINOCCHIO (with 2LPT displacements), COLA (the version by 
\citet{izard2016}), PATCHY, EZmocks and Halogen, plus an implementation
of the lognormal model of \citet{coles1991}.

\begin{figure}
\centering{
\includegraphics[width=0.49\textwidth]{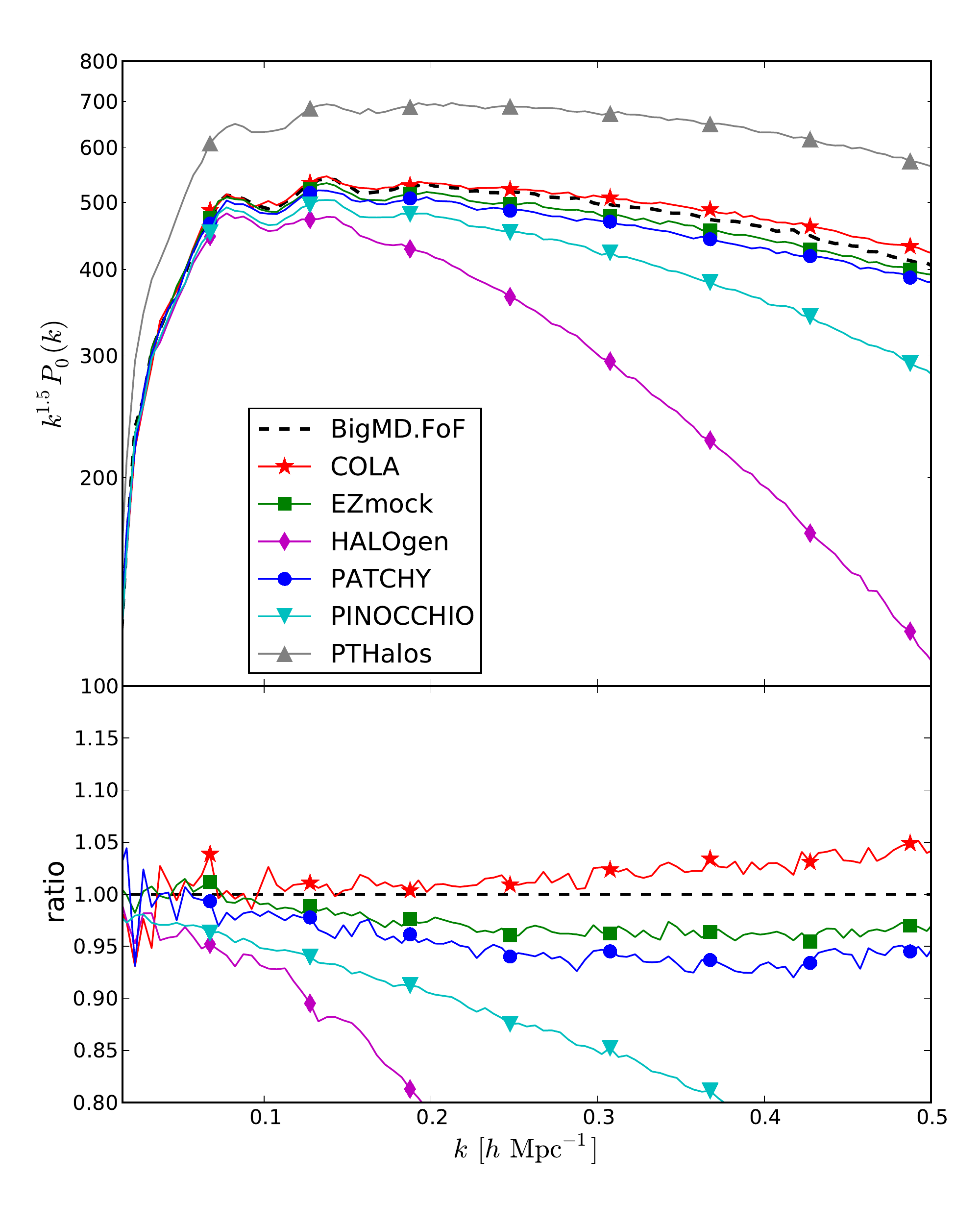}
\includegraphics[width=0.49\textwidth]{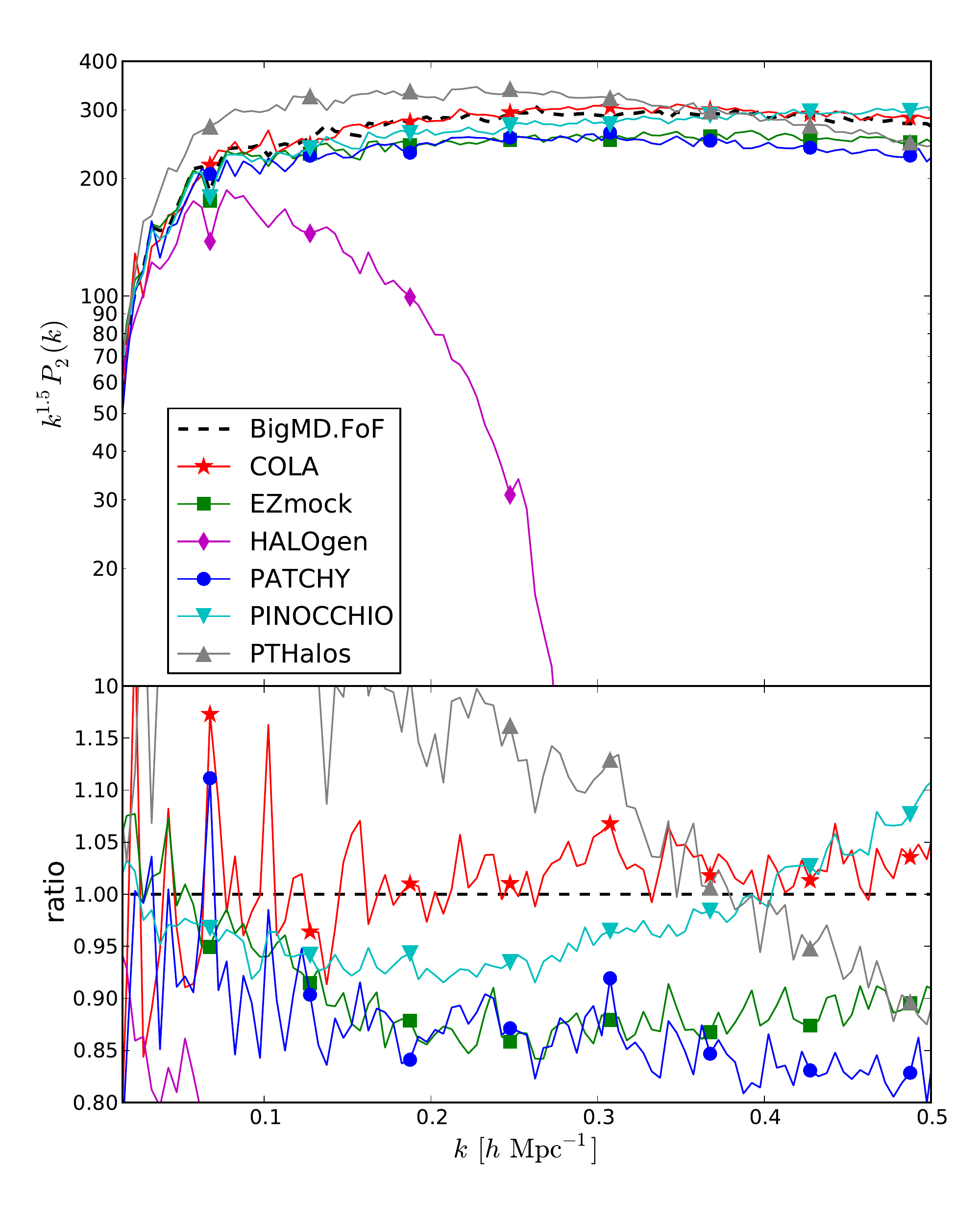}
}
\caption{Comparison of methods for the monopole (left) and the
  quadrupole (right) of the power spectrum in redshift space. Dashed
  lines correspond to the BigMD FoF reference catalog, the other lines
  give the results for the various methods (see the legend). Taken
  from Chuang et al., 2015, {\mnras}, pag. 694, Fig. 7. Published by
  Oxford University Press.}
\label{fig:nifty}
\end{figure}

Tested statistics were two-point correlation function in
real (configuration) and redshift space (monopole and quadrupole),
three-point correlation function in real space (for triangles with two
fixed sides), power spectrum in real (Fourier) and redshift (monopole
and quadrupole) space, bispectrum in real space (again for triangles
with two fixed sides). Everything was done at one redshift and one
mass cut, no result found at other redshifts made it worth to inflate
the number of figures beyond the already generous level.
We report here (Figure~\ref{fig:nifty}) figure 7 of 
\citet{nifty}, where the monopole and quadrupole of the redshift-space
power spectrum of halos are shown.

The broad conclusion from that paper is that many of these models are
able to follow the clustering of halos up to third order, to within an
accuracy of $\sim5-10$\% for two-point statistics, though small scales
are a challenge for some of the methods. The exceptions were PTHalos,
that in this implementation was predicting a high bias, Halogen, that
was found to need more development to reach the maturity of the other
codes, and the lognormal model, that was giving wrong results at the
three-point level. The other codes all give good results in
configuration space, performance in Fourier space is more variable.
PINOCCHIO, based on 2LPT displacements in that case, produces a power
spectrum that drops by 10\% at $k=0.2$ \hMpc, as expected. Remarkably,
this lack of power does not affect the quadrupole. The same problem is
in principle shared by PATCHY (to a higher $k$ due to augmentation)
and even more by EZmocks, where the ZA mass density drops below 10\%
already at $k=0.1$ \hMpc. But these two methods are able to correct
for this loss of power with their bias models; bias will then
overweight small-scale density fluctuations to reach the right amount
of power up to three-point statistics. Conversely, the smart N-body
solver COLA shows a very good degree of predictive power, being always
among the best performing methods.

\begin{figure}
\centering
\includegraphics[width=15cm]{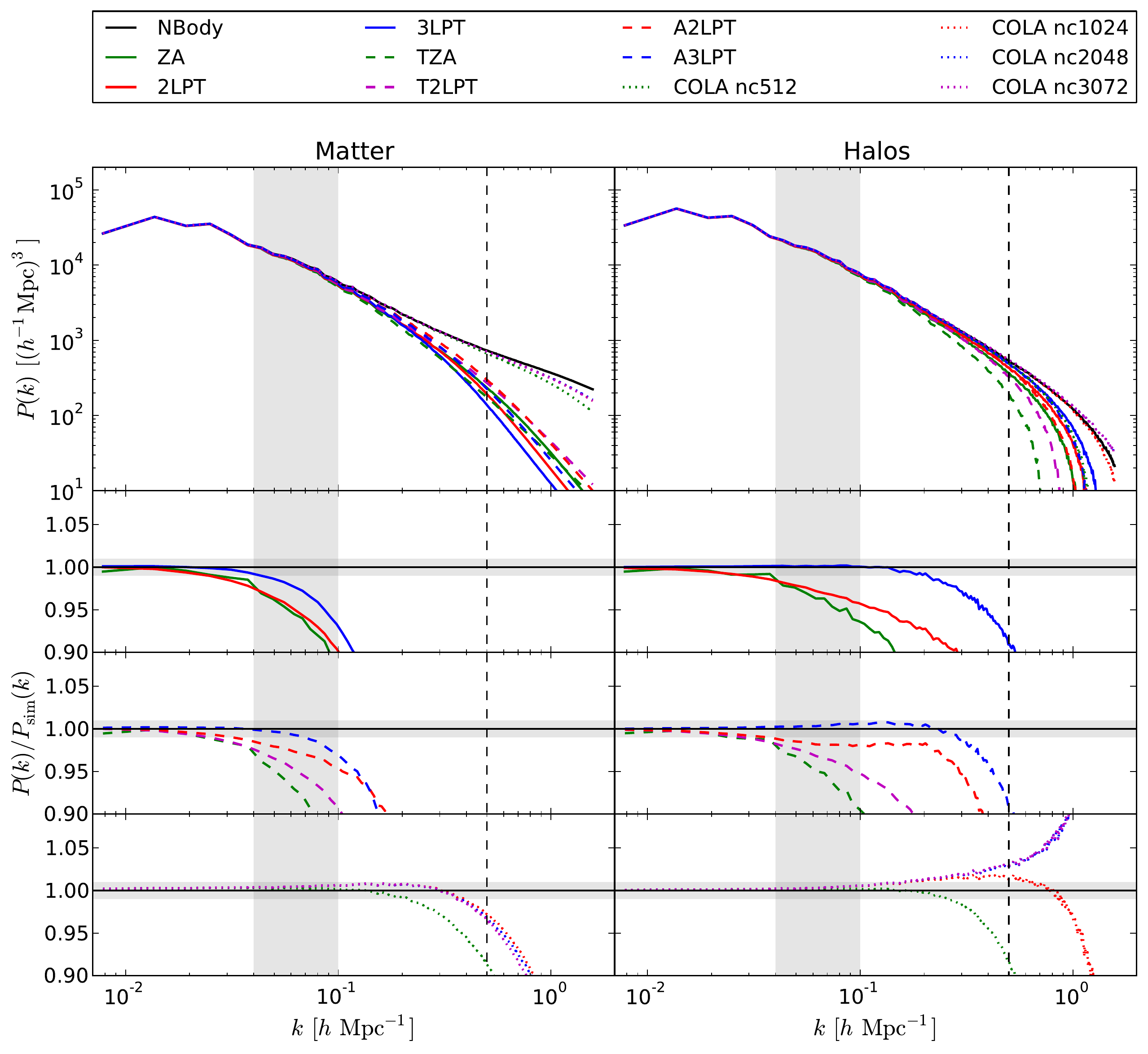}
\caption{Power spectrum of matter and halo density for various methods
  to displace particles, with particle assignment to halos given by
  the simulation. Left and right panels give respectively the matter
  and halo power spectra. The upper panels report all the methods
  given in the legend above the figure. The other panels give
  residuals for the three LPT orders, for the truncated and augmented
  versions, and for the COLA implementations. { Shaded vertical
    areas mark the region of the first BAO, while shaded horizontal areas in
    the residual panels give the region of $\pm1$ \%. }The vertical
  line at $k=0.5$ {\hMpc} marks the scales where the 1-halo term
  becomes dominant.}
\label{fig:munari}
\end{figure}   

\subsection{Generating displacement fields of halos without worrying
  about particle assignments to halos}

Approximate methods of the Lagrangian class (as defined above) are, as
a matter of fact, required to solve two problems at the same time. On
the one hand, they must solve the ``Lagrangian'' problem of assigning
particles to halos, possibly with fine time sampling so as to
reconstruct halo merger histories. On the other hand, they must solve
the ``Eulerian'' problem of producing the large-scale field of
displacements and velocities, up to a mildly non-linear scale as near
as possible to that where the 1-halo term becomes dominant, so as to
displace (in real and redshift space) the center of mass of halos at
the right place.

{ Errors due to approximations in the solution of these two problems have a different impact}
on the predictions of clustering.
An error on assignment of particles to
halos will result in a different set of halos selected for a given
mass or number density cut, and because halo bias sensibly depends on
halo mass, it will affect the normalization of the clustering
statistics. Conversely, an inaccuracy on the matter displacement field
will typically produce a cutoff in $k$ beyond which non-linear power
is not properly recovered.

Producing a displacement field of all particles is a step in common
with bias-based methods. In fact, the same requirements on the
accuracy of reproduction of clustering can be applied to the matter
density field as well as to that of dark matter halos. The first is
not affected by bias, but a cutoff in $k$ is produced in both cases.
For instance, LPT is known to be able to reproduce non-linear
clustering up to a wavenumber $k$ of order of a few times 0.1 \hMpc,
after which orbit crossing and the thickening of pancakes act
similarly to a Gaussian smoothing \citep[see, e.g.][]{monaco2013}. We
show below that this fact has a different impact on matter density and
halo clustering.

Asking an approximate method to guess the assignment of a particle to
a halo is a tough requirement. Conversely, the large-scale
displacement and velocity fields of halos are easier to reproduce if a
prior knowledge of particle assignments is given. One can perform this
exercise: take an N-body simulation, run FoF on a snapshot at some
redshift $z$ and store particle assignments to halos. Then run an
approximate method for the displacement of the matter field, and
compute where halos are expected to be (and what peculiar velocity
they are expected to have) assuming knowledge of particle assignments
to halos and computing halo properties as averages over their
particles.

This exercise has recently been carried on by \citet{munari2016b} in a
paper that is presently in preparation. There are two reasons why this
exercise is interesting. The first is that, for any sensible halo
finder, halos encompass multi-stream regions, so averaging over the
Lagrangian patch that ends up into a halo can be considered as an
optimal and very adaptive way of smoothing the initial density field.
This way one can investigate the ability of approximate methods to
reproduce halo displacements in the most favourable case of perfect
particle assignment to halos, thus setting an upper limit to their
performance. The second reason why this is interesting is that at
least one algorithm, PINOCCHIO, has been shown to be very effective in
predicting particle assignments to halos, even on the basis of the
least accurate ZA displacements. So this test gives an upper limit to
the accuracy that a code like PINOCCHIO can reach if displacements are
computed with a specific approximation.

We anticipate here some of the key results of that paper. The starting
point is a simulation of a 1024 {\Mpch} box sampled with $1024^3$
particles, with initial conditions generated with the {\sc 2LPTic} code of
Scoccimarro \cite{crocce2006}. FoF halos with more than 50 particles
are considered. The same generator of the linear density field 
is used to produce
predictions with implementations of LPT at three orders (ZA, 2LPT,
3LPT\footnote{In the 3LPT implementation the rotational mode is
  neglected.}) in three flavours: straight LPT, truncated LPT,
augmented LPT. We also consider COLA predictions, kindly provided by
Koda \cite{koda2016}, applied to a mesh with cell size $2$, $1$, $1/2$
and $1/3$ times the interparticle distance (meshes of $512^3$,
$1024^3$, $2048^3$ and $3072^3$). Halo positions and velocities are
obtained by averaging over the set of particles that are assigned to
that halo.

Figure~\ref{fig:munari} shows the real-space power spectrum, at $z=0$,
of the matter (left panel) and the halo (right panel) density fields.
Results for the simulation are compared with the following models: LPT
up to third order, truncated Zeldovich approximation (TZA) and T2LPT, augmented A2LPT and A3LPT
and the four COLA variants. Panels below the upper ones show residuals
of the approximate power spectra with respect to that of N-body;
models are grouped in separate panels so as to avoid line crowding.
Power spectra have been computed with the code of Sefusatti
\cite{sefusatti2015}, shot noise has been subtracted. We can draw a
number of conclusions from this figure.

(1) For the matter power spectrum, the wavenumber at which power drops
below a $-10$\% level does not overtake 0.2 {\hMpc} for all models,
with the exception of COLA, that can reach 0.8 {\hMpc} for all meshes
but the coarsest one. Moreover, going from lower to higher orders
provides only some modest improvement, and this is true for all LPT
flavours. This quantifies the trend, noticed in Figure~\ref{fig:lss},
that higher LPT orders may give better convergence before orbit
crossing, but they add to the spreading of the multi-stream region.

(2) All approximate methods are able to recover the halo power
spectrum at a higher wavenumber than the matter power spectrum. For
the straight LPT series, power drops below a $-10$\% level at
$k\sim0.13$, $0.3$ and $0.5$ {\hMpc} for the three orders, showing
that the improvement achieved going to higher orders is significant.

(3) No advantage is found by adopting the truncated LPT scheme. As
noticed in Section~\ref{section:tza}, in the original paper of
Coles et al. \cite{coles1993} the advantage of the truncated scheme
was found to be good for a power spectrum of positive or flat slope
$\ge-1$ but marginal for a slope $-2$, that is shallower than the
slope of a $\Lambda$CDM spectrum at $k\sim0.5$.

(4) Augmentation contributes to the improvement of the LPT series.
This trend is more apparent for the halo power spectrum, in which
case the improvement in going from 2LPT to A2LPT is of the same order
of the improvement in going from 2LPT to 3LPT. A3LPT is only marginally better
than 3LPT.

(5) COLA gives percent accurate results for the matter power spectrum
up to $k\sim0.5$ \hMpc, and is 10\% accurate up to $k=0.8$ \hMpc. Very
good results are obtained for halos, where some excess power at the
2\% level is found at $k=0.5$, but 10\% accuracy is achieved up to
$k=1$ \hMpc. Results are stable with the grid dimension, with the
exception of the coarsest grid that gives poor performance; for halos,
finer grids give some excess of power, the $1024^3$ grid gives the
best performance. Clearly, finer grids are needed by COLA to allow
proper identification of halos, but not to achieve better accuracy in
their clustering.

These results demonstrate the potentiality of approximate methods
based on LPT to correctly place DM halos: both 3LPT and A3LPT
displacements are still 10\% accurate up to $k=0.5$ \hMpc, the scale
at which the 1-halo term is becoming dominant, while COLA outperforms
all methods even with a $1024^3$ grid. This is true, of course,
provided that the Lagrangian part of the problem, particle assignments
to halos, is solved with good accuracy.

{ Before going to the conclusions, I want to stress that averaging
  over the Lagrangian patch of a DM halo is already a form of
  smoothing. Recently, \citet{kopp2016} advocated the use of smoothing
  to improve the predictions of one extension of LPT, where their
  smoothing length is equal to the Lagrangian size of the halo. This
  ``truncation'' is necessary to insert the information of the
  Lagrangian size of a halo into an analytic formalism. So their
  ``choose to smooth'' message is not in contrast with our conclusion
  that truncation does not help LPT in this context.}

\section{Concluding remarks}
\label{section:conclusions}

The search for analytic and semi-analytic approximations to
gravitational clustering, more than being pushed out of fashion, has
been boosted by precision cosmology. This is perfectly in line with
the ideas laid down since the late '60s, when the Russian school
started to investigate the formation of large-scale structure, to the '90s, when
analytic approximations were systematically investigated and used to
constrain the cosmological model. Indeed, the ability to understand
the fine details of clustering pays off when the ``blind'' solution of
N-body simulations must be demonstrated to be accurate to a certain
level, or must be replicated a very large number of times to quantify
the systematics of a galaxy survey.

One of the reasons why approximate methods are acceptable, that was
pointed out by Scoccimarro and Sheth in the PTHalos presentation paper
\cite{scoccimarro2002}, is that the very non-linear baryonic processes
that lead to the formation of galaxies drastically limit the
possibility to accurately predict what happens on small scales,
roughly where the one-halo term becomes dominant. On larger scales,
uncertainties in galaxy formation mainly emerge as an uncertainty in
galaxy bias, that as mentioned in Section~\ref{section:recent} can be
directly measured from the data using higher-order correlation and
marginalized over when estimating cosmological parameters.

This review is mostly concerned with the interpretation of galaxy
clustering. Galaxy weak lensing is another, very powerful tool for the
clustering of matter, that has the great advantage of directly
testing the gravitational potential along a line of sight. Galaxies
here are not unfaithful tracers of the large-scale density field, but
just background objects whose ellipticity is perturbed by
gravitational lensing. The issue in this case is whether galaxy
ellipticities are correlated, and to properly subtract this nuisance
term. There is in principle no reason why approximate methods cannot
be used to generate realizations of matter and gravitational potential
distributions. But if the use of galaxy clustering is limited to the
range of scales that are dominated by the 2-halo term, $k\sim0.5$
{\hMpc}, galaxy lensing has the power to directly investigate the
potential down to much smaller scales, at least to $k\sim10$ \hMpc.
Here the details of halo structure, as ellipticity or substructure,
become determinant. This issue was addressed by \citet{pace2015}, who
computed both the matter power spectrum and the lensing potential of
an N-body simulation trying different experiments: take only particles
found in halos or outside halos; use only halos to compute the density
field; rotate halos to randomize their ellipticities; make halos
isotropic by rotating their particles to erase substructure and
ellipticity; displace halos with 2LPT. Remarkably, the removal of
substructure by isotropization of halos was found to have a small
impact on $P(k)$ up to $0.5$ \hMpc, but to influence the lensing
angular power spectrum at $l > 300$. Clearly, the assumption of
spherical and smooth halos done by all mock generators does not allow
to reach good accuracy. A way out may be to construct halo models that
reproduce ellipticities and substructure, as done with the Matter
density distributiOn Kode for gravitationAl lenses (MOKA) of
\citet{giocoli2012}, that was developed to make predictions on strong
lensing by galaxy clusters.

Throughout this paper I described how to produce catalogs of DM halos,
but this is only the first step in the construction of a mock galaxy
catalog. DM halos must be populated with galaxies, that should follow
a given luminosity function, so that a flux cut can be applied to the
galaxy distribution to produce a realistic mock. This step can be
performed using three different techniques. Halo occupation
distribution models
\cite[e.g.][]{berlind2002,yang2003,skibba2009,zehavi2011,crocce2015}
populate each DM halo with a number of galaxies that depends on its
mass, and this dependence is obtained by requiring to reproduce the
number density (or luminosity function in some cases) and correlation
function of galaxies. In subhalo abundance matching models
\cite[e.g.][]{vale2004,conroy2006} the simulation is able to resolve
substructures of DM halos, named sub-halos, and these are populated
with galaxies assuming a one-to-one correspondence. The third class is
that of SAMs of galaxy formation
\cite{baugh2006,benson2010,somerville2015}, applied to the merger
trees of an N-body simulation \cite[one example for
  all,][]{merson2013}. Because the exact correspondence of galaxies
and halos or sub-halos depends on the complex physics of galaxy
formation, there can be no guarantee of accuracy in this step.
Cosmological measurements will be highly accurate as long as they are
not influenced much by it.



\vspace{6pt} 


\acknowledgments{ I warmly thank Emiliano Munari for letting me
  publish Figures~\ref{fig:lss} and \ref{fig:munari} while our paper
  is in preparation. I have taken deep profit from being part of large
  collaboration, in particular the Euclid Consortium, where I
  constantly interact with many of the people involved in this field.
  The nIFTy comparison project was another stimulating environment to
  understand the potentialities of the various mock generators. I want
  to thank Stefano Borgani, Chia-Hsun Chuang, Francisco-Shu Kitaura,
  Jun Koda, Florent Leclercq, Aseem Paranjape, Cristiano Porciani,
  Ariel Sanchez, Emiliano Sefusatti, Francisco Villaescusa-Navarro and
  Gustavo Yepes for many discussions and for their comments on the
  draft. The paper was improved thanks to the constructive comments of three
  competent referees. The author has been supported by the program
  ``Finanziamento di Ateneo per progetti di ricerca scientifica - FRA
  2015'' of the University of Trieste.

}



\abbreviations{The following abbreviations are used in this manuscript:\\

\noindent 
BAO: baryonic acoustic oscillations\\
CDM: cold dark matter\\
DM: dark matter\\
EPT: Eulerian Perturbation Theory\\
FFT: fast Fourier transform\\
FoF: friends-of-friends\\
LPT: Lagrangian Perturbation Theory\\
2LPT: 2nd-order LPT\\
3LPT: 3rd-order LPT\\
ALPT: augmented LPT\\
PDF: probability distribution function\\
SAM: semi-analytic model\\
ZA: Zeldovich approximation\\
(model acronyms are not reported here)
}



\bibliographystyle{mdpi}


\bibliography{review}


\end{document}